\documentclass[twocolumn,pra,aps,amsmath,amssymb,superscriptaddress,showkeys,10pt]{revtex4-2}

\usepackage{graphicx}
\usepackage[colorlinks=true,citecolor=blue]{hyperref}
\usepackage{braket}
\usepackage{lipsum}
\usepackage{tocbasic}

\usepackage{upgreek,mathrsfs}
\newcommand{\tx}{\text}
\newcommand{\fig}[1]{Fig.~\ref{fig:#1}}
\newcommand{\eq}[1]{Eq.~(\ref{eq:#1})}
\newcommand{\Sec}[1]{Section~\ref{sec:#1}}

\newcommand{\gm}{\gamma}

\newcounter{oldtocdepth}

\newcommand{\hidefromtoc}{%
  \setcounter{oldtocdepth}{\value{tocdepth}}%
  \addtocontents{toc}{\protect\setcounter{tocdepth}{-10}}%
}

\newcommand{\unhidefromtoc}{%
  \addtocontents{toc}{\protect\setcounter{tocdepth}{\value{oldtocdepth}}}%
}

\newcommand{\gcorrm}{5}
\newcommand{\momfromg}{7}
\newcommand{\simplL}{9}
\newcommand{\recursive}{10}
\newcommand{\lamax}{11}
\newcommand{\critone}{15}

\newcommand{\critthree}{20}
\newcommand{\critfour}{21}
\newcommand{\critthreeimproved}{22}
\newcommand{\critfive}{23}
\newcommand{\critfiveimproved}{24}

\begin{document}

\title{Higher-Order Photon Statistics\\ as a New Tool to Reveal Hidden Excited States in a Plasmonic Cavity}
 
\author{Philipp Stegmann}\email{psteg@mit.edu}
\affiliation{Department of Chemistry, Massachusetts Institute of Technology, Cambridge, Massachusetts 02139, USA}

\author{Satyendra Nath Gupta}
\affiliation{Department of Chemical and Biological Physics, Weizmann Institute of Science, Rehovot 761001, Israel\looseness=-1}
\affiliation{Physical Research Laboratory, Ahmedabad 380009, India}

\author{Gilad Haran}
\affiliation{Department of Chemical and Biological Physics, Weizmann Institute of Science, Rehovot 761001, Israel\looseness=-1}

\author{Jianshu Cao}
\affiliation{Department of Chemistry, Massachusetts Institute of Technology, Cambridge, Massachusetts 02139, USA}

\date{\today}

\begin{abstract}
Among the best known quantities obtainable from photon correlation measurements are the $g^{(m)}$~correlation functions. Here, we introduce a new procedure to evaluate these correlation functions based on higher-order factorial cumulants $C_{\tx{F},m}$ which integrate over the time dependence of the correlation functions, i.e., summarize the available information at different time spans. In a systematic manner, the information content of higher-order correlation functions as well as the distribution of photon waiting times is taken into account. Our procedure greatly enhances the sensitivity for probing correlations and, moreover, is robust against a limited counting efficiency and time resolution in experiment. It can be applied even in case $g^{(m)}$ is not accessible at short time spans. We use the new evaluation scheme to analyze the photon emission of a plasmonic cavity coupled to a single quantum dot. We derive criteria which must hold if the system can be described by a generic Jaynes-Cummings model. A violation of the criteria can be explained by the presence of an additional excited quantum dot state.
\end{abstract}

\keywords{photon statistics, correlation functions, plasmonic cavities, quantum dots, excitons, factorial cumulants}

\maketitle

\hidefromtoc
\section{Introduction}
Photon correlation functions are an important concept in the analysis of quantum optical systems~\cite{glauber_quantum_1963}. The nature of the emitted light can be characterized conveniently by the second-order correlation function~$g^{(2)}(t)$~\cite{loudon_quantum_2000}. Classical description leads to $g^{(2)}(0)>1$ for chaotic or thermal sources emitting photons in bunches~\cite{kondakci_sub-thermal_2016,van_charge_2019,brange_photon_2019}. In contrast, quantum emitters as single atoms or molecules~\cite{bamba_origin_2011,nair_biexciton_2011,xu_full_2013,vester_photon_2015,Blazquez_enhancing_2017,huang_nonreciprocal_2018,schaeverbeke_single_2019} can give rise to $g^{(2)}(0)<1$, indicating nonclassical light in form of an antibunched photon stream. Moreover, the value $g^{(2)}(0)$ can be used to verify the presence of quantum coherence~\cite{sanchez_photon_2020} or to obtain the number of quantum emitters~\cite{sykora_exploring_2007,Chung_extracting_2007,amgar_higher-order_2019,loudon_quantum_2000}.  

Higher-order correlation functions $g^{(m)}$ provide a more precise characterization of photon statistics, which is a crucial requirement in the development of quantum technologies~\cite{OBrien_photonic_2009} and the engineering of single-photon emitters~\cite{Aharonovich_solid_2016}. The third-order correlation function can be used to differentiate between the conventional and unconventional photon blockade effect~\cite{Radulaski_photon_2017,huang_nonreciprocal_2018,You_reconfigurable_2020}.
Higher-order antibunching ($g^{(m)}(0)<0$) can be identified~\cite{klyshko_nonclassical_1996,stevens_third-order_2014,qi_multiphoton_2018} even though in second order the photon stream is bunched~\cite{stevens_third-order_2014,rundquist_nonclassical_2014, hamsen_two-photon_2017,bin_two-photon_2018,qi_multiphoton_2018}.
Universal relations between the different correlation functions have also been found for certain system classes~\cite{klyshko_nonclassical_1996,amgar_higher-order_2019}. 

The analysis of correlation functions is typically focused on the limit of vanishing time delay $t=0$. Though, finite-time correlations are measured as well, their interpretation is more challenging~\cite{loudon_quantum_2000} and the complexity increases with each order.
In this paper, we introduce a new procedure to evaluate the correlation functions systematically at finite times. The procedure is based on factorial cumulants $C_{\tx{F},m}$~\cite{beenakker_counting_2001,kambly_factorial_2011,stegmann_detection_2015,stegmann_short_2016} and addresses certain experimental and theoretical issues. First, the evaluation is robust against a limited time resolution. Second, the counting efficiency of the measurement apparatus does not lead to systematic errors, i.e., errors with a nonzero mean. A small efficiency can be compensated by increasing the measurement time and acquiring more data. Third, photon emitters can be distinguished systematically even in case $g^{(m)}$ is not accessible at short time delays. Third, we combine the information from the correlation functions with the concept of waiting times ~\cite{carmichael_photoelectron_1989,cao_generic_2008} which provides complementary statistical insight.

\begin{figure}[t]
	\includegraphics[width=0.48\textwidth]{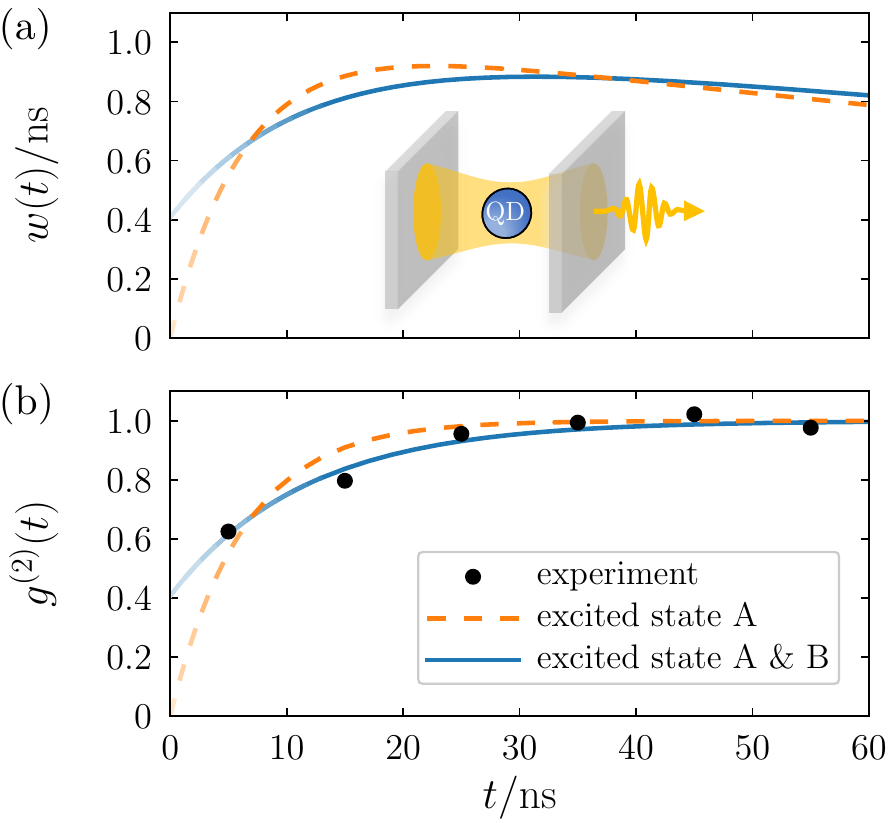}
	\caption{Photon emission statistics of a plasmonic cavity coupled to a single quantum dot. (a)~Waiting-time distribution between consecutive photon counts. (b) Second-order correlation function. The experimental signal is binned in time intervals $\Delta t=10\, \tx{ns}$ (black dots). The parameters used for the simulation (solid blue and dashed orange curves) are given in the Methods section.\vspace{-0.07cm}}\label{fig:fig1}
\end{figure}

The waiting-time distribution $w$ between successive photon emission events can be constructed from all orders of the correlation functions $g^{(m)}$, but is also directly measurable in experiment~\cite{vyas_photon_1989,verberk_photon_2003,delteil_observation_2014}. The distribution characterizes photon bunching and antibunching in a more precise manner than the second-order correlation function~\cite{zhang_quantum_2018,brange_photon_2019,avriller_photon_2021}. Nevertheless, the difference between both quantities is small for times smaller than the average waiting time. Waiting times have been used frequently to study the statistics of electron currents in nanoscale junctions~\cite{brandes_2008,rajabi_waiting_2013,haack_distributions_2014,sothmann_electronic_2014,kosov_waiting_2017,engelhardt_tuning_2019,rudge_counting_2019,Nathan_counting_2019,Stegmann_electron_2021,davis_electronic_2021,Kleinherbers_synchronized_2021,Stegmann_Statistical_2021,Landi_waiting_2021} and enzymatic reactions~\cite{Lu_single_1998,cao_event_2000,cao_generic_2008,Daniel_analysis_2010,Avila_generic_2017,Piephoff_generic_2018,Kumar_transients_2021}. However, they have been used seldom in the analysis of single-photon emitters~\cite{zhang_quantum_2018,avriller_photon_2021}.

In this paper, we follow the experimental work of Reference~\onlinecite{Gupta_complex_2021} and study the photon emission of a plasmonic cavity with an embedded quantum dot sketched in \fig{fig1}(a).
Details regarding the experimental setup are given in the Methods section.
Plasmonic cavities have been investigated intensively in recent years for their ability to be strongly coupled to quantum emitters~\cite{Schlather_near_2013,torma_strong_2014,Otten_origins_2016,Wersall_observation_2017,hugall_plasmonic_2018,Leng_strong_2018,Yankovich_visualizing_2019,Strong_pelton_2019,Bitton_vacuum_2020,You_reconfigurable_2020,Gupta_complex_2021}. Here, the plasmonic cavity is made of a silver bowtie structure,  and a single quantum dot is inserted in the gap at the center of the bowtie.
Such systems have been described very successfully in literature by the well-known \textit{Jaynes-Cummings model}~\cite{Bruce_Jaynes_1993,Agarwal_quantum_2009}, where the dot is modeled as a two-level system with a ground and excited state~A. The resulting waiting-time distribution $w(t)$ and second-order correlation function $g^{(2)}(t)$ are shown as dashed orange curves in \fig{fig1}.

We will discuss how to identify deviations from this generic type of photon statistics due to the presence of a second excited state~B (see solid blue curves) strongly coupled to the cavity. So the single quantum dot is modeled as three-level system with a ground and two excited states (A \& B).

The lifetime of state B is too short to be resolved in the experiment, which gives rise to nonvanishing values in the short-time limit $t\to 0$. Such a feature is not reproducible if we assume just a single excited state~A.
For weaker temporal resolution, we obtain the experimental data depicted by black dots. The shortest time resolved is $t=5\, \tx{ns}$ where both the dashed orange and the solid blue curve are almost identical. So, one may assume that additional excited states can be neglected in the experiment.
We will demonstrate that the presence of the additional state B can be inferred nevertheless by accumulating the information contained in the photon statistics at all available time spans. Thereby, we will also elaborate on the required time resolution and find a remarkable high critical value of $37 \, \tx{ns}$.

\section{Extended Jaynes-Cummings Model}\label{sec:models}
\subsection*{Full Description}

\begin{figure*}[t]
\center
	\includegraphics[width=1.00\textwidth]{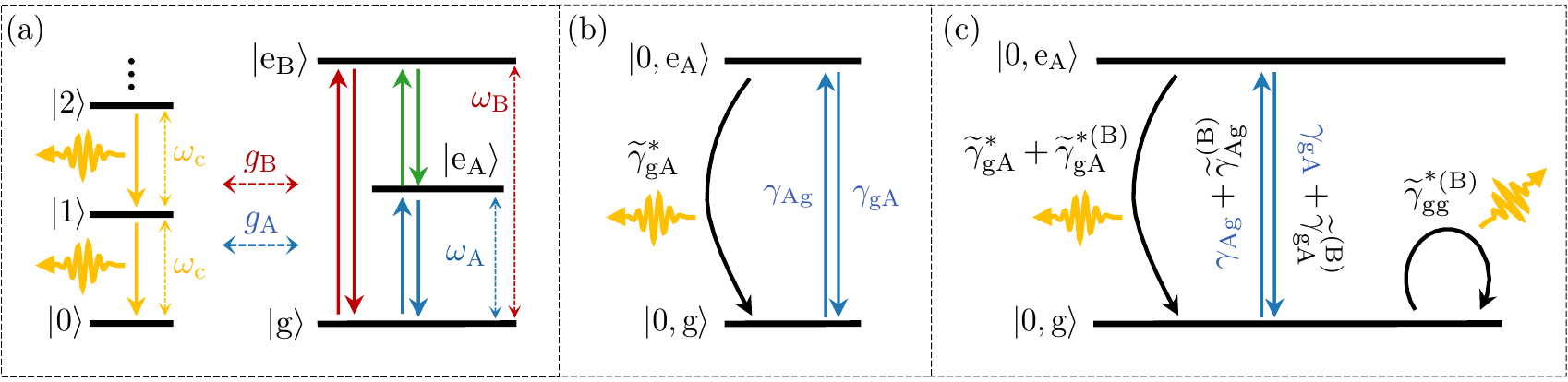}
	\caption{\textcolor{black}{Extended Jaynes-Cummings model: a single-mode cavity coupled to a quantum dot with a ground and two excited states A,B. Incoherent transitions are depicted by solid arrows. (a) Full energy diagram. (b),(c) Effective incoherent model applicable if the cavity field, the excited state B, and coherences decay quickly. The state B is neglected in (b).}}\label{fig:fig2}
\end{figure*}

We study a plasmonic cavity hosting a single quantum dot. The system is described by the extended Jaynes-Cummings Hamiltonian
\begin{equation}\label{eq:extJCH}
 H_{\tx{exJC}}= \omega_\tx{c} a^\dagger  a+ \sum_{i=\tx{A,B}}\left[ \omega_i  \sigma^+_i  \sigma ^-_i  + g_i( a  \sigma^+_i +  a^\dagger  \sigma^-_i)\right],
\end{equation}
where we set $\hbar = 1$ as we do throughout this paper. A diagram of the involved states is depicted in \fig{fig2}(a). The first term describes a single-mode cavity where the operator $a^\dagger$ ($ a$) creates (annihilates) a boson at energy~$\omega_\tx{c}$. The dot is modeled as three-level system with a ground~$\ket{\tx{g}}$ and two excited states $\ket{\tx{e}_\tx{A/B}}$. The lowering and raising operator are $\sigma^-_i = \ket{\tx{g}}\bra{\tx{e}_i}$ and $\sigma^+_i= \ket{\tx{e}_i}\bra{\tx{g}}$. Jaynes-Cummings constants $g_i$ characterize the coupling strengths between the cavity and three-level system. The form of the coupling term requires that the rotation wave approximation holds~\cite{Bruce_Jaynes_1993,Agarwal_quantum_2009}. So, we assume that $\omega_i +\omega_\tx{c}$ is much larger than all other system parameters, in contrast to $\omega_i -\omega_\tx{c}$.

The dynamics of the system are determined by a master equation
\begin{equation}\label{eq:Liofull}
\dot {\rho} = \mathcal{L}  \rho= -i[  H_{\tx{exJC}},  \rho]+ \mathcal{D} \rho \,.
\end{equation}
The first term describes the unitary evolution of the system. The dissipator
\begin{equation}
\mathcal{D}=\kappa \mathcal{D}_{ a}+\hspace{-1mm}\sum_{i=\tx{A,B}}(\gamma_{i\tx{g}} \mathcal{D}_{\!  \sigma^+_i}+\gamma_{\tx{g}i} \mathcal{D}_{\!  \sigma^-_i}+ \gamma_{ii}\mathcal{D}_{\!  \sigma^+_i \sigma^-_i}+\gamma_{\bar i i}\mathcal{D}_{\! \sigma^+_{\bar i} \sigma^-_{i}})\, .
\end{equation}
accounts for incoherent dynamics.
The Lindblad operator reads $\mathcal{D}_{x} \rho = x \rho x^\dagger - \frac{1}{2}\left\{  x^\dagger  x,  \rho \right\}$.
Population transfer from the exciton $i=\tx{A,B}$ to $\bar{i}=\tx{B,A}$ is modeled by $\gamma_{\bar i i}\mathcal{D}_{\! \sigma^+_{\bar i} \sigma^-_{i}}$ and most likely due to phonon-mediated processes~\cite{werschler_coupling_2016}.
Dephasing is taken into account by $\gamma_{ii} \mathcal{D}_{ \sigma^+_i  \sigma^-_i}$.
The term $\gamma_{i\tx{g}} \mathcal{D}_{ \sigma^+}$ describes incoherent pumping. It accounts for pumping of the quantum-dot states via another higher-energy state that is directly excited by an incident monochromatic laser followed by dephasing.
The term $\kappa \mathcal{D}_{ a}$ and $\gamma_{\tx{g}i} \mathcal{D}_{ \sigma^-_i}$ describe the intrinsic decay of an excitation in the cavity and quantum dot~\cite{Novotny_principles_2012}, respectively. The quantum yield of the two kinds of dot excitons outside the cavity is quite different~\cite{Gupta_complex_2021}. So, we can assume a bright and a dark exciton. The intrinsic decay of the latter one is most likely due to heat emission.
\pagebreak

\subsection*{Effective Description}
In this paper, we concentrate on the experimentally relevant parameter regime 
\begin{equation}\label{eq:regime}
\gamma_{i \tx{g}}<\gamma_{\tx{g} i} , \gamma_{i\bar i} \lesssim \Delta t^{-1} \ll g_\tx{A} \ll g_\tx{B},\omega_i,\omega_\tx{c},\gamma_{ii}, \kappa\, ,
\end{equation}
where the state B has only a weak impact on the waiting-time distribution and the second-order correlation function for times $t> 5\,\tx{ns}$ as illustrated in \fig{fig1}.
The photon stream is sampled at a finite time resolution~$\Delta t$. So, the dynamics determined by $ g_i,\omega_i,\omega_\tx{c},\gamma_{ii}$, and $\kappa\,$ are too fast to be resolved in real time. We can eliminate adiabatically the corresponding degrees of freedom, which leads to effective corrections for the slower dynamics. The details of the derivation are presented in the Supporting Information~I. We use the stationary condition~\cite{Cao_Optimization_2009} to remove the fast degrees of freedom, i.e., coherent superpositions, the cavity field as well as the excited state B (due to the strong coupling $g_\tx{B}$). 
We obtain an effective incoherent kinetic rate equation for the states $\ket{0,\tx{g}}$ and $\ket{0,\tx{e}_\tx{A}}$ where $\ket{0}$ is the vacuum state of the cavity. Momentary transient excitations lead to effective incoherent transitions rates~$\widetilde \gamma_{ji}$ between these states, where $i$ is the initial and $j$ the final state.

If we neglect the state B completely, we obtain the dynamics illustrated in \fig{fig2}(b). Pumping of the ground state leads to the excited state~A at rate $\gamma_\tx{Ag}$. The decay of state A at rate $\gamma_\tx{gA}$ is enhanced by the cavity-induced correction $\widetilde \gamma_\tx{gA}^*$, which is well-known as the Purcell effect~\cite{Novotny_principles_2012}.

If we take the adiabatic corrections from state B into account, we obtain the effective dynamics illustrated in \fig{fig2}(c). The excitation rate, the intrinsic decay, and the Purcell enhancement change by $\widetilde \gamma_\tx{Ag}^{(\tx{B})}$, $\widetilde \gamma_\tx{gA}^{(\tx{B})}$, and $\widetilde \gamma_\tx{gA}^{*(\tx{B})}$, respectively.
Moreover, the strong coupling $g_\tx{B}$ leads to the formation of two polaritonic states. However, once populated, the polaritons decay immediately on time scales that can not be resolved in the experiment. The total process of slow excitation and immediate decay can be modeled by the single rate $\widetilde \gamma^{\tx{*(B)}}_\tx{gg}\approx \gamma_\tx{Bg}$ as illustrated by the self-loop in \fig{fig2}(c). This process persists even in the nonresonant case $\omega_\tx{B}\neq \omega_\tx{c}$ if $(\omega_\tx{B}-\omega_\tx{c})^2 \lesssim g_\tx{B}^2(\gamma_\tx{BB}+\kappa)/(\gamma_\tx{AB}+\gamma_\tx{gB})$ (see Supporting Information~I).

In conclusion, the additional state B gives rise to Poissonian photon statistics as long as the system is in its ground state. Such a statistical feature can not be generated in the framework of an ordinary Jaynes-Cummings model with just the single excited state A. We will discuss in the following sections how this contribution to the total photon emission can be identified from the photon-counting statistics.

\section{Photon statistics}\label{sec:FCSphoton}
The photon-counting statistics can be characterized by correlation functions~\cite{glauber_quantum_1963,loudon_quantum_2000}
\begin{equation}\label{eq:gm}
g^{(m)}(t_1,\cdots t_{m}) = \frac{\braket{a^\dagger(t_1) \cdots a^\dagger(t_{m}) a(t_{m})\cdots  a(t_1)}}{\braket{ a^\dagger (t_1) a (t_1)}^m} \,,
\end{equation}
of order $m$ with $t_1<t_2<\cdots <t_m$. They can be calculated from the Liovillian as explained in the Supporting Information II.B.
The correlation functions at zero time delay are related to the occupation number $n$ of the cavity. A seminal expression used frequently is~\cite{loudon_quantum_2000}
\begin{equation}
\braket{n^{(m)}}=\braket{n}^m g^{(m)}(0)\, ,
\end{equation}
with the factorial power $n^{(m)}=n(n-1)\cdots (n-m+1)$. Correlation functions at nonzero time delay have previously received less attention in literature. They are related to the number $N$ of counted photons during a time interval $[0,t]$. We find for the corresponding \textit{factorial moments} (Supporting Information II)
\begin{equation}\label{eq:momfromg}
\braket{N^{(m)}}\! (t)  =   m!  I_\tx{ph}^m \int_0^t \! 	\int_{t_1}^{t} \! \! \cdots \! {\int_{t_{m-1}}^{t}} 
\!\!\! g^{(m)}(t_1,\cdots t_{m})  \tx{d}t_{m}  \! \cdots   \tx{d}t_{1} .
\end{equation}
The mean photon current $I_\tx{ph}=\braket{N}/t= \kappa \braket{n}$ is related to cavity loss rate $\kappa$ and the mean occupation number~$\braket{n}$ of the cavity mode.
From a theory perspective, these moments can be derived conveniently as derivatives $\braket{N^{(m)}}=\partial_z^m \mathcal{M}(z,t)|_{z=1}$ of a generating function~\cite{Plenio_quantum_1998} (Supporting Information II.A)
\begin{equation}\label{eq:gen_maintext}
\mathcal{M}(z,t)= \tx{Tr}\left[ e^{\mathcal{L}_z t} \rho_\tx{NESS} \right].
\end{equation}
The Liouvillian $\mathcal{L}_z=\mathcal{L}_0 +z \mathcal{J}$ with the counting variable $z$ characterizes the system dynamics completely and, thereby, distinguishes between a part increasing the photon counter $\mathcal{J}\rho= \kappa a \rho a^\dagger$ and a part ($\mathcal{L}_0= \mathcal{L} - \mathcal{J}$) leaving the counter unchanged. 
The Liouvillian determining the time evolution of the density matrix is recovered as $\mathcal{L}=\mathcal{L}_1$.
The system is subjected to constant pumping and we assume that counting starts when the non-equilibrium steady state $\rho_\tx{NESS}$ has been reached which fulfills $\mathcal{L}_1 \rho_\tx{NESS}=0$.
For the extended Jaynes-Cummings model depicted in \fig{fig2}(c), the effective Liouvillian (derived in the Supporting Information~I) takes the form
\begin{equation}\label{eq:simplL}
\mathcal{L}_{z} =\!
\begin{pmatrix}
(z\tx{-}1)\,\widetilde \gm^{\tx{*(B)}}_\tx{gg}\tx{\,-\,} \gm_\tx{Ag}\tx{\,-\,}\widetilde \gm_\tx{Ag}^{\tx{(B)}} & \,z (\widetilde\gm^*_\tx{gA}\tx{\,+\,}\widetilde \gm^{\tx{*(B)}}_\tx{gA}) \tx{\,+\,} \gm_\tx{gA} \tx{\,+\,} \widetilde \gm^{\tx{(B)}}_\tx{gA}\\
\gm_\tx{Ag}\tx{\,+\,}\widetilde \gm_\tx{Ag}^{\tx{(B)}}  & \tx{-\,}\widetilde\gm^*_\tx{gA}\tx{\,-\,}\widetilde \gm^{\tx{*(B)}}_\tx{gA} \tx{\,-\,} \gm_\tx{gA} \tx{\,-\,} \widetilde \gm^{\tx{(B)}}_\tx{gA}
\end{pmatrix}
\end{equation}
in the basis $(\rho_{\ket{0,\tx{g}}},\rho_{\ket{0,\tx{e}_\tx{A}}} )^T$.

The generating function defined in~\eq{gen_maintext} can be simplified in the limit of long time intervals $t$. It takes on the large deviation form $\mathcal{M}(z,t) \propto  e^{\lambda_\tx{max}(z)t}$~\cite{Touchette_large_2009}. Here, $\lambda_\tx{max}(z)$ is the eigenvalue of the Liouvillian with the largest real part. This motivates the definition of factorial cumulants $C_{\tx{F},m}(t)=\partial^m_z \ln \mathcal{M}(z,t)|_{z=1}$
which are related to the moments via the recursive formula
\begin{equation}\label{eq:recursive}
C_{\tx{F},m}(t)=\braket{N^{(m)}}(t)+\sum_{i=1}^{m-1} {{m-1}\choose{i-1}} C_{\tx{F},i}(t)    \braket{N^{(m-i)}}(t)
\end{equation}
and give access to the different derivatives of the dominating eigenvalue
\begin{equation}\label{eq:lamax}
\partial_z^m\lambda_\tx{max}(z)|_{z=1}=\lim_{t\to \infty} \frac{C_{\tx{F},m}(t)}{t}\, .
\end{equation}
This information can be used to reconstruct the characteristic polynomial of an unknown Liouvillian, a scheme dubbed \textit{inverse counting statistics}~\cite{bruderer_inverse_2014,stegmann_inverse_2017}. The Liouvillian of the extended Jaynes-Cummings model given in \eq{simplL} has a characteristic polynomial of the form
\begin{equation}\label{eq:charpolgen}
\chi (z,\lambda)=\lambda^2+(a_{01}+a_{11}z) \lambda+ a_{00}+a_{10} z\, ,
\end{equation}
with the coefficients related to the system parameters by
\begin{subequations} 
\begin{align}
a_{11}&=-\widetilde \gm^{\tx{*(B)}}_\tx{gg} , \label{eq:a11} \\
a_{01}&=\gm_\tx{Ag}+\widetilde \gm_\tx{Ag}^{\tx{(B)}}+\widetilde\gm^*_\tx{gA}+\widetilde \gm^{\tx{*(B)}}_\tx{gA} + \gm_\tx{gA} + \widetilde \gm^{\tx{(B)}}_\tx{gA} , \\
\begin{split}
a_{00}&=\left(\gm_\tx{gA}+ \widetilde \gm_\tx{gA}^{\tx{(B)}} \right)  \widetilde \gm_\tx{gg}^{*\tx{(B)}} \\
&\phantom{=}+\left(\widetilde\gm^*_\tx{gA}+\widetilde \gm^{\tx{*(B)}}_\tx{gA}  \right) \left( \gm_\tx{Ag} +\widetilde \gm_\tx{Ag}^{\tx{(B)}} + \widetilde \gm_\tx{gg}^{*\tx{(B)}}\right).
\end{split}
\end{align}
\end{subequations}
Moreover, we have $a_{10}=-a_{00}$ since the existence of a non-equilibrium steady state demands that $\lambda=0$ is an eigenvalue for $z=1$.

The coefficients can be obtained from measured cumulants. We take advantage of \eq{lamax}, which allows us to express the derivatives $\partial_z^m \chi(z,\lambda_\tx{max}(z))_{z=1}=0$ as a function of the coefficients and measured cumulants.
The first three derivatives form a set of linear equations that we solve for the coefficients. As a result, we obtain
\begin{equation}\label{eq:a11viaC}
a_{11}=\lim_{t\to \infty}\left( \frac{3 C_{\tx{F},2}^2 C_{\tx{F},1}}{(2C_{\tx{F},3}C_{\tx{F},1}-3C_{\tx{F},2}^2)t} -\frac{C_{\tx{F},1}}{t} \right)
\end{equation}
and similar expressions for $a_{01}$ and ${a_{00}}$. 
The coefficients $a_{01}$ and $a_{00}$ depend on the dynamics of both excited states A and B. In contrast, $a_{11}$ depends only on the rate~$\widetilde \gm^{\tx{*(B)}}_\tx{gg}$ which vanishes in the absence of the excited state~B. Setting the left-hand side of \eq{a11viaC} to zero, we find an asymptotic condition that must be fulfilled for the third cumulant
\begin{equation}\label{eq:C3C2}
\widetilde{C}_{\tx{F},3}=\frac{3C_{\tx{F},2}^2}{C_{\tx{F},1}}
\end{equation}
in the long-time limit.

Violation of this relation indicates the presence of the state B. We have modified the symbol of the third cumulant by a tilde $\widetilde{C}_{3}$ to indicate that the expression holds in general only for two state kinetics. We emphasize that the factorial cumulants can be expressed by the $g^{(m)}$ correlation functions as shown in the Supporting Information II.D; therefore, \eq{C3C2} is a relation between the correlation functions. Moreover, the extended James-Cummings model determined by the effective Liovillian in \eq{simplL} is a renewal system which is in its ground state after a photon is detected. Thus, the third-order correlation function takes the factorized form $g^{(3)}(t_1,t_2,t_3)=g^{(2)}(t_2-t_1)g^{(2)}(t_3-t_2)$, and only $g^{(2)}$ is required to evaluate \eq{C3C2}.

\begin{figure*}[t]
	\includegraphics[width=0.98\textwidth]{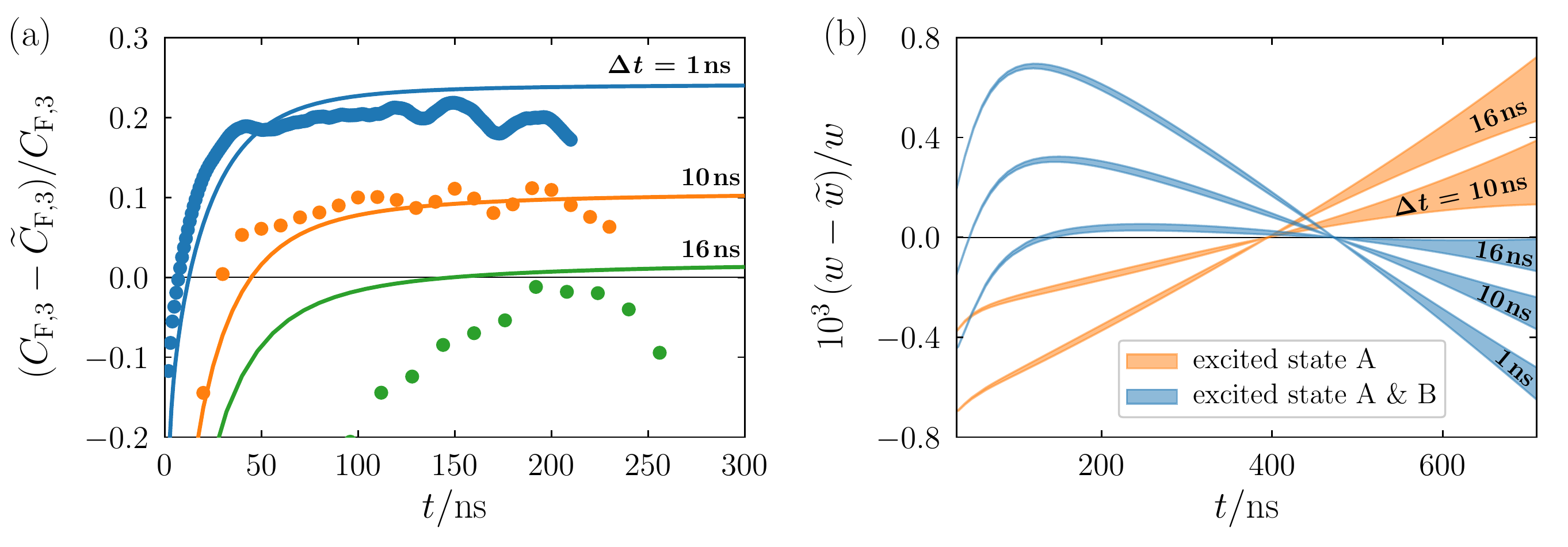}
	\caption{Testing for the presence of an additional state B using (a) the third factorial cumulant and (b) the waiting-time distribution at different sampling times $\Delta t$. The criteria given in Eqs.~(\ref{eq:criw}) and (\ref{eq:criw}) are evaluated. In the long-time limit, a positive sign in (a) and a negative sign in (b) indicate the state B. A sampling time faster than $17\, \tx{ns}$ is required. Dots in (a) are measured in the experiment, whereas solid curves are simulations. Continuous error bars in (b) are simulated assuming that the photon stream is recorded for $100\,\tx{s}$ with $\eta=1$. Other parameters are given in the Methods section.}\label{fig:fig3}
\end{figure*}

\section{Waiting-time distribution}

A quantity providing complementary information to the correlation functions is the waiting-time distribution~$w(t)$~\cite{brandes_2008,rajabi_waiting_2013,haack_distributions_2014,sothmann_electronic_2014,kosov_waiting_2017,engelhardt_tuning_2019,rudge_counting_2019,Nathan_counting_2019,Stegmann_electron_2021,davis_electronic_2021,Kleinherbers_synchronized_2021,zhang_quantum_2018,avriller_photon_2021}. It is the probability density that two consecutive photons are detected at the time difference~$t$. The distribution can be expressed as $w(t)=\braket{\tau}\partial_t^2 P_0(t)$, where $\braket{\tau}=1/I_\tx{ph}$ is the mean waiting time and $P_0=\mathcal{M}(0,t)$ is the idle-time probability that no photons have been counted during the time span $[0,t]$~\cite{Albert_electron_2012}.

The idle-time probability is related to the dominating eigenvalue by $\lambda_\tx{max}(0)=\lim_{t\to \infty} [\ln P_0 (t)]/t$. Therefore, we have an additional relation to obtain the coefficients of the characteristic polynomial. We solve the set of linear equations $\chi(z,\lambda_\tx{max}(z))_{z=0}=0$, $\partial_z \chi(z,\lambda_\tx{max}(z))_{z=1}=0$, and $\partial_z ^2\chi(z,\lambda_\tx{max}(z))_{z=1}=0$ which leads to
\begin{equation}
a_{11} = \lim_{t \to \infty} \left( \frac{C_{\tx{F},2} (\ln P_0)^2 - 2 C_{\tx{F},1}^2\ln P_0 - 2 C_{\tx{F},1}^3  }{(C_{\tx{F},2} \ln P_0 + 2 C_{\tx{F},1} \ln P_0 + 2 C_{\tx{F},1}^2 )t} \right)\, .
\end{equation}
If the state~B is absent and $a_{11}$ vanishes, we obtain the condition
\begin{equation}
\widetilde P_0= \exp \left[ \frac{C_{\tx{F},1}^2}{C_{\tx{F},2}}\left(1-\sqrt{1+2\frac{C_{\tx{F},2}}{C_{\tx{F},1}}} \right) \right]
\end{equation}
and for the waiting-time distribution
\begin{equation}\label{eq:criwait}
\widetilde w = \braket{\tau} \widetilde P_0 \left(\frac{\ln \widetilde P_0}{t}\right)^2 
\end{equation}
in the long-time limit.
Detecting a violation of the relations in \eq{C3C2} and (\ref{eq:criwait}) is a sensitive way to identify the presence of the state B. However, one must ensure that the violation is not caused by measurement imperfections which we discuss in detail in the following sections. 

\section{Photon detection}\label{sec:detect}
Single-photon detection is always subjected to experimental limitations. Intensity losses lead to a finite probability~$\eta$ that an emitted photon is detected. Moreover, the continuous photon stream is sampled at finite time resolution~$\Delta t$ which leads to a discrete-time signal of $t/\Delta t$ points.
The original generating function given in \eq{gen_maintext} must be adjusted accordingly
\begin{equation}\label{eq:Mdetect}
\mathcal{M}(z,t) =  \tx{Tr}\biggl(\biggl \{\sum_{i=0}^\infty\frac{\Delta t^i}{i!}\bigl[\eta \mathcal{L}_z+ (1-\eta ) \mathcal{L}_1 \bigr]^i_{z^n \to z}   \biggr \}^{t/\Delta t}\hspace{-2mm}  \rho_\tx{NESS}\biggr).
\end{equation}
During the sampling time $\Delta t$, a maximum of one photon can be detected. If more photons reach the detector (counting factor $z^n$ with $n > 1$), the detector counter increases only by one (counting factor $z$).
In our experimental setup, emitted photons pass a 50/50 beam splitter before being focused on two single-photon detectors, each recording its own time trace. So, two photons can be detected during each time interval $[t-\Delta t/2, t+\Delta t/2]$. In our simulations [single-detector setup, \eq{Mdetect}], we set $\Delta t \to \Delta t/2$ to allow for the detection of two photons during these time intervals. Similarly, the numeric values for $\Delta t$ given in the following must be divided by a factor of 2 to reproduce the results in experiment by a single-detector setup.

Naturally, the relations given in \eq{C3C2} and (\ref{eq:criwait}) do not hold exactly due to the modified generating function. The relative deviation from the predicted behavior of the third factorial cumulant [\eq{C3C2}] is illustrated in \fig{fig3}(a) for the sampling times $\Delta t=1\, \tx{ns}, 10\, \tx{ns}$, and $16\, \tx{ns}$. Simulations by means of \eq{Mdetect} are depicted by solid curves and experimental data by dots. The parameters are given in the caption and the Methods section.

\begin{figure*}[t]
	\includegraphics[width=0.98\textwidth]{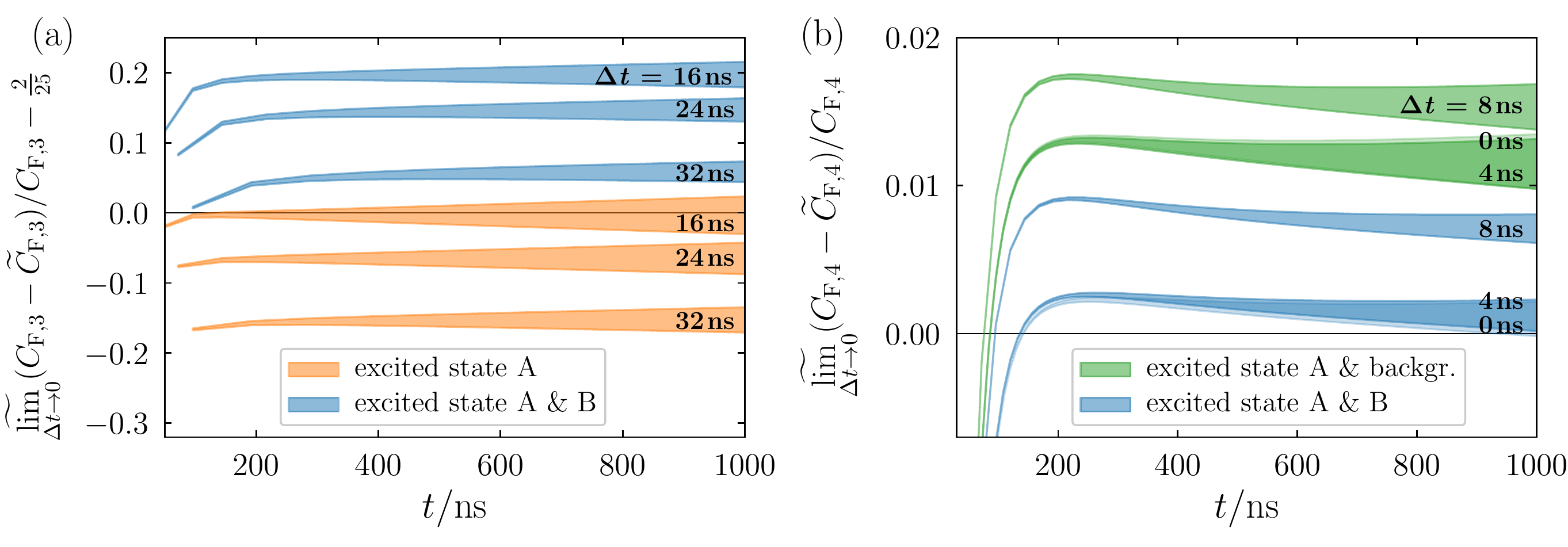}
	\caption{Testing for the presence of an additional state B using (a) the third and (b) fourth factorial cumulant while compensating for the finite sampling time $\Delta t$ during photon detection, i.e., the limit $\Delta t\to 0$ is approximated by a Newton series evaluating the data at $\Delta t$, $2\Delta t$, and $3\Delta t$. Continuous error bars for the expressions given in Eqs.~(\ref{eq:crit3improv}) and (\ref{eq:crit4improv}) are depicted. (a) The state B yields a positive sign for $\Delta t < 37\, \tx{ns}$. (b) A vanishing value in the long-time limit indicates state~B for $\Delta t < 5\, \tx{ns}$, whereas background photons yield a nonvanishing value. We assumed that the photon stream is recorded for $500\,\tx{s}$ in (a) and for $10^7\tx{s}$ in (b) with $\eta=1$. Other parameters are given in the Methods section.}\label{fig:fig4}
\end{figure*}

If we neglect state B, the relative deviations in the long-time limit take the form
\begin{equation}\label{eq:crit3}
\frac{C_{\tx{F},3}-\widetilde{C}_{\tx{F},3}}{C_{\tx{F},3}}= 1-\frac{3(\Delta t'^2-2 e^{-\Delta t'}+2)^2}{2\Delta t'^2(\Delta t'^2+6)}<0 \, ,
\end{equation}
with $\Delta t'=(\gm_\tx{gA}+\widetilde\gm^*_\tx{gA}) \Delta t$. In general, the relative deviation is nonvanishing. However, its sign is fixed, irrespective of the system or detector parameters. Therefore, an opposite sign reveals that the dynamics can not be modeled just by a single excited state. Unfortunately, the presence of the additional state B does not guarantee a different sign (Supporting Information~III). The sampling time $\Delta t$ must be smaller than a certain critical value, which for the parameters used in \fig{fig3} is about $17\,\tx{ns}$. Otherwise, the state B can not be resolved anymore.

The criterion given in \eq{crit3} and the $g^{(2)}$-correlation function do not depend on the detection probability~$\eta$ of the emitted photons, i.e., no systematic error is introduced. Decreasing the detection probability in the experiment leads only to more statistical noise which can be compensated by increasing the measurement time and acquiring more data.
In contrast, the waiting-time distribution is much more affected by a finite detection probability. We present simulations of the relative deviation from \eq{criwait} in \fig{fig3}(b). We assume that the photon stream is recorded for 100\,\tx{s} which yields the depicted continuous error bars~\cite{kleinherbers_pushing_2022} (Supporting Information~IV). In the long-time limit, the deviations take the form
\begin{equation}\label{eq:criw}
\frac{w-\widetilde{w}}{w}= - t \frac{(\Delta t'^2+6)(\eta \, \gm_{\tx{Ag}}\, \widetilde\gm^*_\tx{gA})^3}{3(\gm_{\tx{gA}}+\widetilde\gm^*_\tx{gA})^5} \frac{C_{\tx{F},3}-\widetilde{C}_{\tx{F},3}}{C_{\tx{F},3}} > 0 .
\end{equation}
Again a different sign indicates the presence of the state~B. However, deviations decrease with $\eta^3$. Thus, a small detection probability makes it much more challenging to identify a certain sign in case of statistical noise due to a limited amount of data. 

To surpass the time resolution limit, we suggest the following procedure which comes at the cost of low statistical noise. First, the recorded discrete-time signal is downsampled by a factor of two and three. We obtain photon streams at sampling time $2\Delta t$ and $3\Delta t$. Then, we evaluate the relative error $(C_{\tx{F},3}-\widetilde C_{\tx{F},3})/C_{\tx{F},3}$ for each stream separately.
We assume that the error caused by the finite time resolution can be expanded in a Newton series around $2\Delta t$.
The expansion has the form $f(x)=f(2\Delta t) + f'(2\Delta t)(x-2\Delta t)+ \frac{f''(2\Delta t)}{2}(x-2\Delta t)^2$ with $f'(2\Delta t)=[f(3\Delta t)-f(\Delta t)]/(2\Delta t)$ and $f''(2\Delta t)=[f(3\Delta t)-2f(2\Delta t)+f(\Delta t)]/\Delta t^2$.
From this expansion, we can approximate the result that would be obtainable for vanishing sampling time. Finally, the criterion in \eq{crit3} takes the modified form
\begin{equation}\label{eq:crit3improv}
\underset{\Delta t \to 0}{  \widetilde \lim}  \left( \frac{C_{\tx{F},3}-\widetilde{C}_{\tx{F},3}}{C_{\tx{F},3}} \right)  < \frac{2}{25}\, ,
\end{equation}
with $\widetilde \lim_{\Delta t \to 0} f(\Delta t):=3 f(\Delta t) -3 f(2\Delta t) +f(3\Delta t) $.
The maximal value of the error is $2/25$, irrespective of the system and detector parameters. This is a consequence of \eq{crit3} depending only on the effective sampling time~$\Delta t'$. The criterion in \eq{crit3improv} is illustrated in \fig{fig4}(a). The critical sampling time has increased significantly to $\,37\,\tx{ns}$.

\begin{table*}[t]
	\begin{tabular}{|c||c|c|c|c|c|c|c|c|c|c|c|c|c|c|c|}
	\hline System
	\begin{minipage}[c][8mm]{0mm}\end{minipage} & \large$\frac{\hbar \gm_\tx{Ag}}{\tx{neV}}$ &\large $\frac{\hbar \gm_\tx{Bg}}{\tx{neV}}$ &\large $\frac{\hbar \gm_\tx{gA}}{\tx{neV}}$ &\large $\frac{\hbar \gm_\tx{gB}}{\tx{neV}}$ &\large $\frac{\hbar \gm_\tx{AB}^0}{\tx{neV}}$ & \large $\frac{\hbar}{\Delta t \, \tx{neV}}$ &\large $\frac{\hbar g_\tx{A}}{\tx{$\mu$eV}}$ &\large $\frac{\hbar g_\tx{B}}{\tx{eV}}$ &\large $\frac{\hbar \omega_\tx{c}}{\tx{eV}}$ &\large $\frac{\hbar \omega_\tx{A}}{\tx{eV}}$ &\large $\frac{\hbar \omega_\tx{B}}{\tx{eV}}$ &\large $\frac{\hbar \gm_\tx{AA}}{\tx{eV}}$ &\large $\frac{\hbar \gm_\tx{BB}}{\tx{eV}}$ &\large $\frac{\hbar \kappa}{\tx{eV}}$ & \large $\frac{\hbar \gm_\tx{noise}}{\tx{neV}}$\\
	\hline\hline
	excited state A & $5.0$ & - & $32.5$ & - & - & $40-600$ & $88.0$ & - & $1.93$ & $1.95$ & - & $0.05$ & - & $0.40$ & -\\
	\hline
	excited state A \& B & $5.0$ & $ 1.1 $  & $32.5$ & $100$ & $100$ & $40-600$ & $20.0$ & $0.10$ & $1.93$ & $1.95$ & $2.00$ & $0.05$ & $0.13$  & $0.40$ & - \\
	\hline
	excited state A \& background  & $5.0$ & - & $32.5$ & - & - & $40-600$ & $47.0$ & - & $1.93$ & $1.95$ & - & $0.05$ & - & $0.40$ &0.5\\
	\hline
	\end{tabular}

\caption{\label{tab1}
System parameters used for the different simulations.
}
\end{table*}

Processing the waiting-time distribution in a similar manner is not practical. The limit $\widetilde \lim_{\Delta t \to 0}(w-\widetilde{w})/w$ scales with $(\eta \, \gm_{\tx{Ag}}\, \widetilde\gm^*_\tx{gA})^3/(\gm_{\tx{gA}}+\widetilde\gm^*_\tx{gA})^5$ and, thus, has no universal bound. The scaling factor must be known if we want to identify the presence of state B.

\section{Background Photons}\label{sec:backgr}
Measurements of photon statistics suffer from background photons which lead to false detector counts. The resulting statistical features are similar to those generated by the case of an additional state B. Especially, the second-order correlation function is nonvanishing in the short-time limit~\cite{brouri_photon_2000}. However, higher-order correctors can be used to identify qualitative differences. 

We incorporate the background photons in our simulation as state-independent contribution to the Liouvillian $ \mathcal{L}_z + (z-1)\gamma_\tx{noise}\mathbf{1}$. False counts are occurring at the rate $\gamma_\tx{noise}$. Then, the characteristic polynomial given in \eq{charpolgen} acquires the additional coefficient $a_{20}$. If the state B is present and the background can be neglected, the coefficient~$a_{20}$ vanishes and we find that the fourth factorial cumulant fulfills
\begin{equation}\label{eq:tCs4}
\widetilde{C}_{\tx{F},4}=\frac{4}{3}\frac{C_{\tx{F},3}^2}{C_{\tx{F},2}} + 2\frac{C_{\tx{F},3}C_{\tx{F},2}}{C_{\tx{F},1}}-3\frac{C_{\tx{F},2}^3}{C_{\tx{F},1}^2}
\end{equation}
in the long-time limit. The details of the derivation are presented in the Supporting Information~V.
The relative deviation from \eq{tCs4} does not have a universal sign for arbitrary system parameters.
Therefore, we have to inspect deviations from zero which requires the measurement errors to be negligible. The error induced by the finite detection probability $\eta$ is suppressed at least by a factor $\mathcal{O}(\Delta t^4)$. To compensate for the leading orders in $\Delta t$, we suggest to analyze the criterion
\begin{equation}\label{eq:crit4improv}
\underset{\Delta t \to 0}{  \widetilde \lim}  \left(\frac{C_{\tx{F},4}-\widetilde{C}_{\tx{F},4}}{C_{\tx{F},4}} \right)  = 0 \, ,
\end{equation}
which holds in the absence of a photon background and for small sampling times. We illustrate the criterion in \fig{fig4}(b). The critical sampling time is $5\, \tx{ns}$. For smaller sampling times, the finite value reveals the presence of background noise.

\section{Conclusions}
We have proposed a new procedure to analyze the information contained in the higher-order correlation functions~$g^{(m)}$ at nonzero time delay. By integrating out time dependences in $g^{(m)}$, we obtain factorial cumulants which give access to features in the photon statistics beyond the generic interpretation schemes of photon correlation functions at zero time delay.

Instead of assuming a specific stochastic model and comparing simulated with measured factorial cumulants, the procedure deals with the opposite problem: if only a few factorial cumulants are given, what can we learn about the underlying stochastic model?
As output of the procedure, we obtain sign criteria whose violations exclude certain candidates for the underlying model. The sign criteria are robust against a limited counting efficiency and time resolution in experiment, i.e., information can be obtained even in case that $g^{(m)}$ is not known at short times. Moreover, it can be used to evaluate the waiting-time distribution which is known to be an information source complementary to correlation functions.

We have demonstrated the procedure for the photon emission of a plasmonic cavity coupled to a quantum dot. We have derived a criterion whose violation indicates that the underlying model is not the ordinary Jaynes-Cummings model. The violation can be explained by modeling the quantum dot as three level system or by including background photons. We have derived a criterion allowing one to distinguishing between both cases.

\section{Methods}
\subsection*{Experimental Setup}
Silver bowtie cavities were prepared as discussed in Reference~\onlinecite{Gupta_complex_2021}. In brief, bowtie cavities were made on SiN grids using electron-beam lithography followed by electron beam silver evaporation and liftoff. Semiconductor quantum dots (CdSe/ZnS) were inserted into the cavities using the capillary force method~\cite{Gupta_complex_2021}. Second-order correlation measurements were performed using a MircoTime 200 (PicoQuant) single-particle spectrometer with a 488 nm laser.

\subsection*{Model Parameters}\label{app:par}
The simulations presented in this paper can be obtained either from the full or the effective description introduced in \Sec{models}. The model parameters for the full description are given in Tab.~\ref{tab1} and guided by~\cite{Gupta_complex_2021}. We assume that population transfer from the state A to state B and vice versa is thermally activated such that
\begin{align}
\gm_{\tx{AB}} &= [ 1+ f_\tx{B} (\hbar \omega_\tx{B} - \hbar \omega_\tx{A}, T)] \gm_{\tx{AB}}^0 \, ,\\
\gm_{\tx{BA}} &= f_\tx{B} (\hbar \omega_\tx{B} - \hbar \omega_\tx{A}, T) \gm_{\tx{AB}}^0 \, ,
\end{align}   
where $f_\tx{B}(E,T)$ is the Bose-Einstein distribution at temperature $T$ (we assume $T= 300\, \tx{K}$) and energy $E$.

We obtain the parameters for the effective description from Tab.~\ref{tab1} following the derivation shown in the Supporting Information I. We obtain $\gm_\tx{Ag}\tx{\,+\,}\widetilde \gm_\tx{Ag}^{\tx{(B)}}=5.0\, \tx{neV}$ and $\gm_\tx{gA} \tx{\,+\,} \widetilde \gm^{\tx{(B)}}_\tx{gA}=32.5\, \tx{neV}$.
If state B and background photons are neglected, we get $\widetilde\gm^*_\tx{gA}=68.3\, \, \tx{neV}$. Taking state B into account yields $\widetilde\gm^*_\tx{gA}+\widetilde \gm^{\tx{*(B)}}_\tx{gA}=19.3\, \, \tx{neV}$ and $\widetilde \gm^{\tx{*(B)}}_\tx{gg}=1.1\, \tx{neV}$. If state B is neglected but significant background photons are taken into account, see \fig{fig4}(b), we have $\widetilde\gm^*_\tx{gA}=19.3\, \, \tx{neV}$ and $\gamma_\tx{noise}=0.5\, \tx{neV}$.

\section{Supporting Information}
Derivation of the effective Liouvillian; Relations between factorial moments, factorial cumulants, and $g^{(m)}$ correlation functions; Analytic expressions for Eqs.~(\ref{eq:crit3}) and (\ref{eq:criw}) in case of an additional state~B or background photons; Details on the calculation of continuous error bars; Inverse counting statistics in case of a general two-state model

\begin{acknowledgments}
We thank E. Kleinherbers for useful discussions. This work was supported by the NSF (CHE 1800301 and CHE 1836913) and the School of Science fund (Sloan Fund) at MIT.
P. S. acknowledges support from the German National Academy of Sciences Leopoldina (Grant No.~LPDS 2019-10). Part of the work was completed during J. Cao's visit to the Weizmann Institute of Science under the sponsorship of the Rosi and Max Varon visiting Professorship.
\end{acknowledgments}

%

\unhidefromtoc
\makeatletter
\renewcommand{\theequation}{S\arabic{equation}}
\renewcommand{\thefigure}{S\arabic{figure}}
\renewcommand{\bibnumfmt}[1]{[S#1]}
\renewcommand{\citenumfont}[1]{S#1}


\newcommand{\stoptocentries}{\renewcommand{\addcontentsline}[3]{}}
\newcommand{\starttocentries}{\let\addcontentsline\oldaddcontentsline}

\newcommand\entrynumberwithdot[1]{#1{.} \hfill}
\renewcommand{\thepart}{}

\makeatletter

\newcommand*{\skillmon@section@dotfill}{
    \def\@dotsep{5.0}\TOCLineLeaderFill}
\DeclareTOCStyleEntry[
 entrynumberformat={\entrynumberwithdot}
,entryformat={\bfseries}
,pagenumberformat=\bfseries
,beforeskip=30pt plus .2pt
    ,numwidth=2.7em,
    ,dynnumwidth=true,
    ,numsep=20em,
   ,linefill=\skillmon@section@dotfill
    ,indent=1.3em
]{tocline}{section}
\DeclareTOCStyleEntry[
 entrynumberformat={\entrynumberwithdot},
 entryformat=\bfseries,
    ,beforeskip=5pt plus .2pt
 ,numwidth=1.8em
    ,linefill=\skillmon@section@dotfill
]{tocline}{subsection}
\DeclareTOCStyleEntry[
 ,entrynumberformat={\entrynumberwithdot}
 ,entryformat={\textit},
    ,beforeskip=5pt plus .1pt
 ,numwidth=1.4em
    ,linefill=\skillmon@section@dotfill
]{tocline}{subsubsection}
\DeclareTOCStyleEntry[
  level=\sectiontocdepth,
  beforeskip=0pt plus .2pt,
  indent=0pt,
  numwidth=1.5em,
  entryformat=\usekomafont{sectionentry}\textsl
]{tocline}{nsection}
\DeclareTOCStyleEntry[
pagenumberformat={\normalsize \bfseries},
 ,entrynumberformat={\hidden}
 ,entryformat={\textbf},
    ,beforeskip=30pt plus .1pt
 ,numwidth=0em
    ,linefill=\skillmon@section@dotfill
    ,indent=4.0em
]{tocline}{part}

\makeatother

\onecolumngrid
\newpage

\fontsize{10}{12}\selectfont

\thispagestyle{empty}
\begin{center}
\textmd{\large Supporting Information for}\\
\vspace{0.5cm}
{\bf \large Higher-Order Photon Statistics\\
as a New Tool to Reveal Hidden Excited States in a Plasmonic Cavity\\}
\vspace{0.5cm}
Philipp Stegmann,$^{1,\,\textcolor{red}{*}}$ Satyendra Nath Gupta,$^{2,3}$ Gilad Haran,$^2$ and Jianshu Cao$^{1}$\\
\vspace{0.1cm}
\textit{\small $^1$Department of Chemistry, Massachusetts Institute of Technology, Cambridge, Massachusetts 02139, USA}\\
\textit{\small $^2$Department of Chemical and Biological Physics, Weizmann Institute of Science, Rehovot 761001, Israel\looseness=-1}\\
\textit{\small $^3$Physical Research Laboratory, Ahmedabad 380009, India}\\
\small (Dated: \today)
\vspace{0.5cm}
\end{center}

{
  \hypersetup{linkcolor=black}
  \vspace{0cm}
  \tableofcontents
}
\vfill
\urlstyle{rm}
\hspace{-0.48cm} \footnotesize * \url{psteg@mit.edu}

\newpage
\setcounter{section}{0}
\setcounter{equation}{0}
\setcounter{figure}{0}
\setcounter{table}{0}
\setcounter{page}{1}

\fontsize{10}{12}\selectfont
\section{EFFECTIVE LIOUVILLIAN}\label{app:effective}
We concentrate on the parameter regime
\begin{equation}\label{eq:regime}
\gamma_{i \tx{g}}<\gamma_{\tx{g} i} , \gamma_{i\bar i} \lesssim \Delta t^{-1} \ll g_\tx{A} \ll g_\tx{B},\omega_i,\omega_\tx{c},\gamma_{ii}, \kappa\, .
\end{equation}
The cavity field, the excited state B, and coherences~\footnote{In case $\gamma_{ii}, \kappa\ll g_\tx{B}$, dephasing is not strong enough to destroy all coherences quickly. However, the coherent dynamics happen at time scales $\propto g_\tx{B}$ too fast to be resolved in experiment. So, all coherences can still be replaced by an effective coarse grained average, i.e., the effective incoherent description of~Section~\ref{app:effective}.} decay quickly.
The Liouvillian becomes effectively two-dimensional. Only the states $\ket{0,\tx{g}}$ and $\ket{0,\tx{e}_\tx{A}}$ are populated where $\ket{0}$ is the vacuum state of the cavity mode. To obtain the effective Liouvillian, we can not simply set the other density matrix elements to zero. Instead, we must redirect probability currents correctly. We follow Ref.~\onlinecite{Cao_Optimization_2009} and use the stationary condition. We solve $0=\bra{i}\mathcal{L}_z \rho_z \ket{j}$ for the density matrix element $\bra{i} \rho_z \ket{j}$ which we want to exclude. In this way, we express $\bra{i} \rho_z \ket{j}$ by all remaining density matrix elements. We insert this expression back into the original master equation and, by doing so, have replaced $\bra{i} \rho_z \ket{j}$ by effective transition rates. We apply this procedure successively starting with the full Liouvillian $\mathcal{L}_z=\mathcal{L} + (z-1) \mathcal{J}$, where $\mathcal{L}$ is given in Eq.~(2) of the main text and the jumping superoperator reads $\mathcal{J} \rho=\kappa  a  \rho  a^\dagger$. The cavity is either empty $\ket{0}$ or singly occupied $\ket{1}$. The details of the derivation are presented in the following. First, we exclude off-diagonal density matrix elements and obtain
\begin{equation}
\mathcal{L}_z=\left(
\begin{array}{cccccc}
\tx{-}\,\gamma_\tx{Ag}\,\tx{-}\,\gamma_\tx{Bg} & \text{$\gamma_\tx{gA}$} & \text{$\gamma_\tx{gB}$} & z\,\kappa   & 0 & 0  \\
 \text{$\gamma_\tx{Ag}$} & \tx{-}\,\gamma_\tx{gA}\,\tx{-}\,\gamma_\tx{BA}\,\tx{-}\,\widetilde{\gamma}_\tx{BA}\,\tx{-}\,\widetilde{\gamma}_\tx{1g,0A} & \text{$\gamma_\tx{AB}$}\,\tx{+}\,\widetilde{\gamma}_\tx{AB}  & \widetilde{\gamma}_\tx{0A,1g} & z\,\kappa   & 0  \\
 \text{$\gamma_\tx{Bg}$} & \text{$\gamma_\tx{BA}$}\,\tx{+}\,\widetilde{\gamma}_\tx{BA} & \text{-$\gamma_\tx{gB}$\,\tx{-}\,$\gamma_\tx{AB}$}\,\tx{-}\,\widetilde{\gamma}_\tx{AB} & \widetilde{\gamma}_\tx{0B,1g} & 0 & z\,\kappa    \\
 0 & \widetilde{\gamma}_\tx{1g,0A}  & \widetilde{\gamma}_\tx{1g,0B}  & \tx{-}\, \kappa \,\tx{-}\, \gamma_\tx{Ag} \,\tx{-}\,\gamma_\tx{Bg}\,\tx{-}\,\widetilde{\gamma}_\tx{0A,1g}\,\tx{-}\,\widetilde{\gamma}_\tx{0B,1g}   & \text{$\gamma_\tx{gA}$} & \text{$\gamma_\tx{gB}$}  \\
 0 & 0 & 0 & \text{$\gamma_\tx{Ag}$} & \tx{-}\,\kappa \,\tx{-}\,\gamma_\tx{gA} \,\tx{-}\, \gamma_\tx{BA}  & \text{$\gamma_\tx{AB}$} \\
 0 & 0 & 0 & \text{$\gamma_\tx{Bg}$} & \text{$\gamma_\tx{BA}$} & \tx{-}\,\kappa \, \tx{-}\,\gamma_\tx{gB} \, \tx{-}\, \gamma_\tx{AB}   
\end{array}
\right)
\end{equation}
in the basis $(\rho_{\ket{0,\tx{g}}},\rho_{\ket{0,\tx{e}_\tx{A}}} ,\rho_{\ket{0,\tx{e}_\tx{B}}} ,\rho_{\ket{1,\tx{g}}} \rho_{\ket{1,\tx{e}_\tx{A}}}, \rho_{\ket{1,\tx{e}_\tx{B}}})^T$. The effective transition rates read 
\begin{align}
\widetilde{\gamma}_{ij}&=2 g_j^2 f_{i j}(g_i^2)\, ,\\
\widetilde{\gamma}_{0i,\tx{1g}}&=\widetilde{\gamma}_{\tx{1g},0i}=2 g_i^2 f_{\bar{i} i}\left\{g_i^2-g_{\bar i}^2+ \left[\gamma^\tx{dph}_\tx{AB}+i(\omega_{\bar{i}}-\omega_i)\right]\left[\gamma^\tx{dph}_{\bar{i}\tx{g}}+i (\omega_{\bar{i}}-\omega_\tx{c})\right] \right\}\,,
\end{align}
with the function

\begin{equation}
f_{ij}(x)=\tx{Re}\, (x / \left\{ g_i^2 \left[ \gamma^\tx{dph}_{i\tx{g}}+i(\omega_i -\omega_\tx{c}) \right]+ \left[g_j^2+\left(\gamma^\tx{dph}_\tx{AB}+i(\omega_i-\omega_j)\right)	\left( \gamma^\tx{dph}_{i\tx{g}}+i(\omega_i -\omega_\tx{c})\right)\left(\gamma^\tx{dph}_{j\tx{g}}+i(\omega_\tx{c} -\omega_j)\right) \right] \right\})
\end{equation}
and $i=\tx{A,B}$ as well as $\bar{i}=\tx{B,A}$. The rates $\gamma^\tx{dph}_\tx{AB}=\sum_{i=\tx{A,B}}(\gamma_{ii}+\gamma_{i\bar{i}}+\gamma_{\tx{g}i})/2$ and $\gamma^\tx{dph}_{i\tx{g}}=(\gamma_{\bar{i}i}+\gamma_{ii}+\gamma_\tx{Ag}+\gamma_\tx{Bg}+\gamma_{\tx{g}i}+\kappa)/2$ determine the dephasing of coherent superpositions between $\ket{0,\tx{e}_\tx{A}},\ket{0,\tx{e}_\tx{B}}$ and between $\ket{1,\tx{g}},\ket{0,\tx{e}_i}$.

Excitations starting from $\rho_{\ket{1,\tx{g}}}$ and leading to $\rho_{\ket{1,\tx{e}_i}}$ are strongly suppressed since $\gamma_{i\tx{g}}$ is the smallest rate in the system and, especially, much smaller than $\kappa$ and $\widetilde{\gamma}_{0i,1\tx{g}}\propto g_i^2$. So, we can set $\rho_{\ket{1,\tx{e}_i}}=0$ and the Liouvillian becomes
\begin{equation}\label{eq:lio4x4}
\mathcal{L}_z=\left(
\begin{array}{cccccc}
\tx{-}\,\gamma_\tx{Ag}\,\tx{-}\,\gamma_\tx{Bg} & \text{$\gamma_\tx{gA}$} & \text{$\gamma_\tx{gB}$} & z\,\kappa    \\
 \text{$\gamma_\tx{Ag}$} & \tx{-}\,\gamma_\tx{gA}\,\tx{-}\,\gamma_\tx{BA}\,\tx{-}\,\widetilde{\gamma}_\tx{BA}\,\tx{-}\,\widetilde{\gamma}_\tx{1g,0A} & \text{$\gamma_\tx{AB}$} +  \text{$\widetilde \gamma_\tx{AB}$} & \widetilde{\gamma}_\tx{0A,1g}   \\
 \text{$\gamma_\tx{Bg}$} & \text{$\gamma_\tx{BA}$}\,\tx{+}\,\widetilde{\gamma}_\tx{BA} & \text{-$\gamma_\tx{gB}$\,\tx{-}\,$\gamma_\tx{AB}$} \,\tx{-}\, \text{$\widetilde \gamma_\tx{AB}$}\, \tx{-} \, \widetilde{\gamma}_\tx{1g,0B}& \widetilde{\gamma}_\tx{0B,1g}   \\
 0 & \widetilde{\gamma}_\tx{1g,0A}  & \widetilde{\gamma}_\tx{1g,0B}  & \tx{-}\, \kappa \,\tx{-}\,\widetilde{\gamma}_\tx{0A,1g} \,\tx{-}\,\widetilde{\gamma}_\tx{0B,1g}   \\
\end{array}
\right)
\end{equation}
in the basis $(\rho_{\ket{0,\tx{g}}},\rho_{\ket{0,\tx{e}_\tx{A}}} ,\rho_{\ket{0,\tx{e}_\tx{B}}} ,\rho_{\ket{1,\tx{g}}})^T$. The elements $\rho_{\ket{0,\tx{e}_\tx{B}}}$ and $\rho_{\ket{1,\tx{g}}}$ exchange occupation probability quickly via $\widetilde{\gamma}_\tx{1g,0B}\,\rho_{\ket{0,\tx{e}_\tx{B}}}$ and $\widetilde{\gamma}_\tx{0B,1g}\,\rho_{\ket{1,\tx{g}}}$. Moreover, $\rho_{\ket{1,\tx{g}}}$ decays quickly at the cavity loss rate $\kappa$. Thus, we can exclude both $\rho_{\ket{0,\tx{e}_\tx{B}}}$ and $\rho_{\ket{1,\tx{g}}}$ from the Liouvillian and obtain
\begin{equation}\label{eq:simplL}
\mathcal{L}_{z} =\!
\begin{pmatrix}
(z\tx{-}1)\,\widetilde \gm^{\tx{*(B)}}_\tx{gg}\tx{\,-\,} \gm_\tx{Ag}\tx{\,-\,}\widetilde \gm_\tx{Ag}^{\tx{(B)}} & \,z (\widetilde\gm^*_\tx{gA}\tx{\,+\,}\widetilde \gm^{\tx{*(B)}}_\tx{gA}) \tx{\,+\,} \gm_\tx{gA} \tx{\,+\,} \widetilde \gm^{\tx{(B)}}_\tx{gA}\\
\gm_\tx{Ag}\tx{\,+\,}\widetilde \gm_\tx{Ag}^{\tx{(B)}}  & \tx{-\,}\widetilde\gm^*_\tx{gA}\tx{\,-\,}\widetilde \gm^{\tx{*(B)}}_\tx{gA} \tx{\,-\,} \gm_\tx{gA} \tx{\,-\,} \widetilde \gm^{\tx{(B)}}_\tx{gA}
\end{pmatrix}
\end{equation}
in the basis $(\rho_{\ket{0,\tx{g}}},\rho_{\ket{0,\tx{e}_\tx{A}}} )^T$, which is~Eq.~(\simplL) in the main text. The effective transition rates can be expressed by probabilities $p_{i,j}=(\mathcal{L}_{z})_{ij}/(-\mathcal{L}_{z})_{jj}$ that the system changes to state $i$ if it starts in state $j$ with the Liouvillian taken from \eq{lio4x4}. The rates are
\begin{align}
\widetilde\gm^{\tx{*(B)}}_\tx{gg} &=p_\tx{0g,1g}\, p_\tx{1g,0B}\sum_{k=0}^\infty (p_\tx{0B,1g}\, p_\tx{1g,0B})^k\, \gamma_\tx{Bg}= \frac{p_\tx{0g,1g}\, p_\tx{1g,0B}}{1-p_\tx{0g,1g}\, p_\tx{1g,0B}}\, \gamma_\tx{Bg}\,, \\
\widetilde\gamma_\tx{Ag}^{\tx{(B)}} &= (p_\tx{0A,0B}+p_\tx{0A,1g}\,p_\tx{1g,0B})\sum_{k=0}^\infty (p_\tx{0B,1g}\, p_\tx{1g,0B})^k\, \gamma_\tx{Bg}=\frac{p_\tx{0A,0B}+p_\tx{0A,1g}\,p_\tx{1g,0B}}{1-p_\tx{0B,1g}\, p_\tx{1g,0B}}\, \gamma_\tx{Bg}\,, \\
\begin{split}
\widetilde\gamma_\tx{gA}^* +\widetilde\gamma_\tx{gA}^{\tx{*(B)}}&=p_\tx{0g,1g} \sum_{k=0}^\infty (p_\tx{1g,0B}\, p_\tx{0B,1g})^k \,\widetilde{\gamma}_\tx{1g,0A} + p_\tx{0g,1g}\, p_\tx{1g,0B}\sum_{k=0}^\infty (p_\tx{0B,1g}\, p_\tx{1g,0B})^k\,
(\gamma_\tx{BA}+\widetilde \gamma_\tx{BA})\,  \\&= \frac{p_\tx{0g,1g}}{1-p_\tx{1g,0B}\, p_\tx{0B,1g}} \widetilde{\gamma}_\tx{1g,0A} + \frac{p_\tx{0g,1g}\, p_\tx{1g,0B}}{1-p_\tx{0B,1g}\, p_\tx{1g,0B}}(\gamma_\tx{BA}+\widetilde \gamma_\tx{BA})\,,
\end{split}\\
\begin{split}
 \widetilde\gm_\tx{gA}^{\tx{(B)}} &= p_\tx{0g,0B}\, \sum_{k=0}^\infty (p_\tx{0B,1g}\, p_\tx{1g,0B})^k \,(\gamma_\tx{BA}+\widetilde \gamma_\tx{BA})+p_\tx{0g,0B}\, p_\tx{0B,1g} \sum_{k=0}^\infty (p_\tx{1g,0B}\, p_\tx{0B,1g})^k \,\widetilde{\gamma}_\tx{1g,0A}\\
 &=\frac{p_\tx{0g,0B}}{1-p_\tx{0B,1g}\, p_\tx{1g,0B}}(\gamma_\tx{BA}+\widetilde \gamma_\tx{BA})+ \frac{p_\tx{0g,0B}\, p_\tx{0B,1g}}{1-p_\tx{1g,0B}\, p_\tx{0B,1g}}\widetilde{\gamma}_\tx{1g,0A}\,,
 \end{split}
\end{align}
where we distinguished between terms originating from state A alone and those requiring the presence of state B indicated by a superscript (B). Especially, we have $\widetilde\gamma_\tx{gA}^*=p_\tx{0g,1g} \,\widetilde{\gamma}_\tx{1g,0A}$ being independent of state B.
Taking only the leading order corrections in $g_\tx{B},\omega_i,\omega_\tx{c},\gamma_{ii},\kappa$ into account, the rates read
\begin{align}
\widetilde\gm_\tx{gg}^{\tx{*(B)}} &=\gamma_\tx{Bg} -\frac{\gamma_\tx{AB}+\gamma_\tx{gB}}{\kappa}\left( \frac{\widetilde{\gamma}_\tx{0B,1g}+\kappa}{\widetilde{\gamma}_\tx{1g,0B}} \right)\gamma_{\tx{Bg}}, \label{eq:gmtgg} \\
\widetilde\gm_\tx{Ag}^{\tx{(B)}} &= \frac{\gamma_\tx{AB}}{\kappa}\left( \frac{\widetilde{\gamma}_\tx{0B,1g}+\kappa}{\widetilde{\gamma}_\tx{1g,0B}} \right)\gamma_{\tx{Bg}}, \label{eq:gmtAD} \\ 
\widetilde\gamma_\tx{gA}^* +\widetilde\gamma_\tx{gA}^{\tx{*(B)}} &=\widetilde{\gamma}_\tx{1g,0A}+ \gamma_\tx{BA}+\widetilde{\gamma}_\tx{BA}  -\frac{\gamma_\tx{AB}+\gamma_\tx{gB}}{\kappa}\left( \frac{\widetilde{\gamma}_\tx{0B,1g}+\kappa}{\widetilde{\gamma}_\tx{1g,0B}} \right)\gamma_\tx{BA}, \label{eq:gmtstargA}\\
\widetilde\gm_\tx{gA}^{\tx{(B)}} &= \frac{\gamma_\tx{gB}}{\kappa}\left( \frac{\widetilde{\gamma}_\tx{0B,1g}+\kappa}{\widetilde{\gamma}_\tx{1g,0B}} \right)\gamma_{\tx{BA}},  \label{eq:gmtgA} 
\end{align} 
with the simplified expressions
\begin{align}
\widetilde{\gamma}_\tx{1g,0A}+\widetilde{\gamma}_\tx{BA}&= \tx{Re} \frac{4 g^2_\tx{A}}{\gamma_\tx{AA} +\kappa -2 i (\omega_\tx{A}-\omega_\tx{c})+\frac{4g_\tx{B}^2}{\gamma_\tx{AA}+\gamma_\tx{BB}+2i (\omega_\tx{B}-\omega_\tx{A})}},\\
 \frac{\widetilde{\gamma}_\tx{0B,1g}+\kappa}{\widetilde{\gamma}_\tx{1g,0B}} &=  \frac{\rho_{\ket{0,\tx{e}_\tx{B}}}}{\rho_{\ket{1,\tx{g}}}}= 1+\frac{(\omega_\tx{B}+\kappa)^2+4(\omega_\tx{B}-\omega_\tx{c})^2}{ \frac{4g_\tx{B}^2}{\kappa} (\gamma_\tx{BB}+\kappa)}.
\end{align}
The rate $\widetilde\gm_\tx{gg}^{*\tx{(B)}}$ accounts for a momentary change of the system's state during which the cavity emits a photon. Starting from the state $\ket{0,\tx{g}}$, the intermediate state $\ket{0,\tx{e}_\tx{B}}$ (strongly coupled to $\ket{1,\tx{g}}$) is populated at rate $\gamma_\tx{Bg}$. The intermediate states $\ket{0,\tx{e}_\tx{B}}$ and $\ket{1,\tx{g}}$ are depleted by the probability currents $\kappa \, \rho_{\ket{1,\tx{g}}}$, $\gamma_{\tx{AB}} \, \rho_{\ket{0,\tx{e}_\tx{B}}}$, and $\gamma_{\tx{gB}} \, \rho_{\ket{0,\tx{e}_\tx{B}}}$. Only the first one leads to the final state $\ket{0,\tx{g}}$ accompanied by the emission of a cavity photon. The depletion via the latter two currents is therefore subtracted from $\widetilde\gm_\tx{gg}^{*\tx{(B)}}$ via the term $-\frac{(\gamma_{\tx{AB}}+\gamma_{\tx{gB}} ) \rho_{\ket{0,\tx{e}_\tx{B}}}}{\kappa \rho_{\ket{1,\tx{g}}}}{\gamma}_\tx{Bg}$. The form of the other rates $\widetilde\gm_\tx{Ag}^\tx{(B)}$,  $\widetilde\gm^*_\tx{gA}$, $\widetilde\gm^\tx{*(B)}_\tx{gA}$, and $\widetilde\gm_\tx{gA}^\tx{(B)}$ can be explained in a similar manner.

Finally, we inspect each element of the Liouvillian given in \eq{simplL} separately taking the parameter regime of \eq{regime} into account. We neglect terms that are suppressed by $\kappa$ except $\widetilde{\gamma}_\tx{1g,0A}\propto g_\tx{A}^2$. We obtain the two-dimensional Liouvillian 
\begin{equation}\label{eq:Liofullsimple}
\mathcal{L}_z=\begin{pmatrix}
(z-1) \gm_\tx{Bg}-\gm_\tx{Ag} & \,\,\,z \left(\widetilde\gm^*_\tx{gA}+\widetilde \gm^{\tx{*(B)}}_\tx{gA}\right) + \gm_\tx{gA}\\
\gm_\tx{Ag} & -\gm^*_\tx{gA}-\widetilde \gm^{\tx{*(B)}}_\tx{gA} - \gm_\tx{gA}
\end{pmatrix},
\end{equation}
with
\begin{equation} 
\widetilde\gm^*_\tx{gA}+\widetilde\gm^{\tx{*(B)}}_\tx{gA}=\gamma_\tx{BA}+\tx{Re} \frac{4 g^2_\tx{A}}{\gamma_\tx{AA} +\kappa -2 i (\omega_\tx{A}-\omega_\tx{c})+\frac{4g_\tx{B}^2}{\gamma_\tx{AA}+\gamma_\tx{BB}+2i (\omega_\tx{B}-\omega_\tx{A})}}.
\end{equation} 

\subsection{Nonresonant Regime}
The Jaynes-Cummings coupling term defined in Eq~(1) of the main text is suppressed in the strongly nonresonant case $\omega_i-\omega_\tx{c} \gg g_\tx{B},\gamma_{ii},\kappa$. To see to what extend the form of the Liovillian given in \eq{simplL} holds, we expand it up to the leading-order terms in $\omega_i-\omega_\tx{c}$. We obtain

\begin{equation}\label{eq:Liofullsimple}
\mathcal{L}_z=\begin{pmatrix}
-\gm_\tx{Ag} - \frac{\gamma_\tx{AB}\gamma_\tx{Bg}}{\gamma_\tx{AB}+\gamma_\tx{gB}}& \gm_\tx{gA}+ \frac{\gamma_\tx{gB}\gamma_\tx{BA}}{\gamma_\tx{AB}+\gamma_\tx{gB}}\\
\gm_\tx{Ag} + \frac{\gamma_\tx{AB}\gamma_\tx{Bg}}{\gamma_\tx{AB}+\gamma_\tx{gB}} &- \gm_\tx{gA}- \frac{\gamma_\tx{gB}\gamma_\tx{BA}}{\gamma_\tx{AB}+\gamma_\tx{gB}}\end{pmatrix}
+
\begin{pmatrix}
(z-1) \gamma_\tx{Bg} {\widetilde{g}_\tx{B}}+  \frac{\gm_\tx{AB}\gamma_\tx{Bg}  {\widetilde{g}_\tx{B}}}{\gm_\tx{AB} + \gm_\tx{gB}}  &	z\left[\gm_\tx{BA} {\widetilde{g}_\tx{B}} + (\gm_\tx{AB} + \gm_\tx{gB}){\widetilde{g}_\tx{A}} \right]	-	\frac{\gamma_\tx{gB} \gm_\tx{BA} {\widetilde{g}_\tx{B}}}{\gm_\tx{AB} + \gm_\tx{gB}} \\
-\frac{ \gm_\tx{AB} \gamma_\tx{Bg} {\widetilde{g}_\tx{B}}}{\gm_\tx{AB} + \gm_\tx{gB}}  & -\frac{\gamma_\tx{AB} \gm_\tx{BA} {\widetilde{g}_\tx{B}}}{\gm_\tx{AB} + \gm_\tx{gB}} - (\gm_\tx{AB} + \gm_\tx{gB}){\widetilde{g}_\tx{A}}
\end{pmatrix},
\end{equation}
with
\begin{equation}
\widetilde{g}_i=	\frac{g_i^2 	\left(\gm_{ii} +\gm_{\tx{g}i} + \gm_{\bar{i} i} 	+	\gm_\tx{Ag}	+ \gm_\tx{Bg}	+ \kappa\right)	}{(\gm_\tx{AB}	+ \gm_\tx{gB}) (\omega_i	-	\omega_\tx{c})^2} \approx \frac{g_i^2 	\left(\gm_{ii} + \kappa\right)	}{(\gm_\tx{AB}	+ \gm_\tx{gB}) (\omega_i	-	\omega_\tx{c})^2}.
\end{equation}
Both the Purcell effect and the formation of polaritonic states are suppressed deeply in the nonresonant regime. However, these effects are relevant if $\widetilde{g}_i \gtrsim1$. 

\section{RELATIONS BETWEEN FACTORIAL MOMENTS, FACTORIAL CUMULANTS, AND CORRELATION FUNCTIONS~$g^{(m)}$}\label{app:4momcum}

\subsection{Moments of the Full Counting Statistics}\label{app:FCS}
To study the photon counting statistics, we resolve the density matrix in terms of the number $N$ of emitted photons during a time interval $[0,t]$, i.e., $\rho(t)=\sum_N \rho(N,t)$~\cite{Plenio_quantum_1998}. The master equation for the $N$-resolved density matrix takes the form
\begin{equation}\label{eq:nmaster}
\dot {\rho}(N,t) = \mathcal{L}_0 \rho(N,t) + \mathcal{J} \rho(N-1,t) \, ,
\end{equation}
where $\mathcal{J} \rho=\kappa  a  \rho  a^\dagger$ is the jump superoperator of the photon emission and $\mathcal{L}_0=\mathcal{L}- \mathcal{J}$. We trace out the degrees of freedom of the cavity and the three-level system to obtain the photon counting statistics $P_N(t)=\tx{Tr}[\rho(N,t)]$, i.e., the probability that $N$ photons are emitted during $[0,t]$. Instead of solving \eq{nmaster} directly, we use the generating function technique~\cite{Plenio_quantum_1998}. Equation (\ref{eq:nmaster}) is $z$-transformed
\begin{equation}\label{eq:zmaster}
\dot {\rho}(z,t) = \mathcal{L}_z  \rho(z,t) \, ,
\end{equation}
with $\mathcal{L}_z=\mathcal{L}_0+z \mathcal{J}$ and $ \rho(z,t) = \sum_N z^N  \rho(n,t)$. The solution takes the form $ \rho(z,t)=e^{\mathcal{L}_z t}  {\rho}_\tx{NESS}$, where the non-equilibrium steady state fulfills $ \mathcal{L} \rho_\tx{NESS}=0$ and has been reached before counting starts. Finally, we can introduce the generating function
\begin{equation}\label{eq:gen}
\mathcal{M}(z,t)=\sum_N z^N P_N(t)= \tx{Tr}\left[ e^{\mathcal{L}_z t} \rho_\tx{NESS} \right]
\end{equation}
of generalized factorial moments~\cite{stegmann_detection_2015}
\begin{equation}\label{eq:momsm}
M_{\mathfrak{s},m}(t)=\partial_z^m \mathcal{M}(z,t)|_{z=\mathfrak{s}}=\sum_N N^{(m)} \mathfrak{s}^{N-m} P_n(t)\,,
\end{equation}
with the factorial power $N^{(m)}=N(N-1)\cdots (N-m+1)$. We can emphasize different parts of the counting statistics by varying the parameter $\mathfrak{s}$. We obtain factorial moments $\braket{N^{(m)}}:=M_{1,m}=\sum_N N^{(m)} P_N$ \cite{Cao_Correlations_2006} for $\mathfrak{s}=1$ which summarize the information contained in the $P_N$ with $N\ge m$. For $\mathfrak{s}=0$, we obtain the moments $M_{0,m}=m! P_m$ and, therefore, concentrate on the information contained in specific $P_m$.
In general, positive values of $\mathfrak{s}$ have been studied in the context of dynamical ~\cite{Cilluffo_Microscopic_2021,Fodor_irreversibility_2022,Macieszczak_theory_2021,brange_non_2022} and topological phases~\cite{Engelhardt_random_2017,Riwar_fractional_2019}, whereas negative values have turned out to emphasize correlations~\cite{Souto_quench_2017,Kleinherbers_Revealing_2018,stegmann_detection_2015} and coherent dynamics~\cite{Stegmann_Coherent_2018,Droste_finite_2016}.

\pagebreak
\subsection{Correlation Functions~$g^{(m)}$}
The correlation functions can be expressed conveniently by the jump operator and the Liouvillian as illustrated in this subsection.
We start with Eq.~(\gcorrm) of the main text
\begin{equation}\label{eq:gm}
g^{(m)}(t_1,\cdots t_{m}) = \frac{\braket{a^\dagger(t_1) a^\dagger(t_2)\cdots a^\dagger(t_{m}) a(t_{m})\cdots a(t_2) a(t_1)}}{\braket{ a^\dagger (t_1) a (t_1)}^m}\, .
\end{equation}
Knowing the time evolution of the density matrix $\rho(t+\tau)=e^{\mathcal L \tau} \rho(t)$, we can take advantage of the quantum regression theorem~\cite{Gardiner_quantum_2010} and write  
\begin{equation}\label{eq:time}
g^{(m)}(t_1,\cdots t_{m}) = \frac{\tx{Tr}\left[ a 	e^{\mathcal L (t_{m}-t_{m-1}) } 	\left\{a 	e^{\mathcal L (t_{m-1}-t_{m-2})}	\cdots 	\left[a e^{\mathcal L (t_{2}-t_{1})} (a  \,\rho_\tx{NESS} \,	a^\dagger) a^\dagger \right]	\cdots  a^\dagger \right\}	a^\dagger \right]}{ \tx{Tr}[ a  \,\rho_\tx{NESS} \, a^\dagger]^m}\,,
\end{equation}
where we have assumed that the initial system is in its non equilibrium steady state $\rho(t_1)=\rho_\tx{NESS}$.
We make use of the jumping operator notation $\mathcal J \rho= \kappa a \rho a^\dagger$ and obtain
\begin{equation}\label{eq:gmcompact}
g^{(m)}(t_1,\cdots t_{m}) = \frac{1}{I_\tx{ph}^{m}}	\tx{Tr}\left[ \mathcal{J} 	e^{\mathcal L (t_{m}-t_{m-1}) } 	\cdots 	\mathcal{J}  e^{\mathcal L (t_{2}-t_{1})} \mathcal{J} \rho_\tx{NESS}\right]\,,
\end{equation}
with the mean photon current $I_\tx{ph} = \tx{Tr}[\mathcal{J} \rho_\tx{NESS}]$. The correlation functions depend only on $m-1$ time differences $\tau_i=t_{i+1}-t_{i}$. Therefore, we can introduce the compacter notation
\begin{equation}\label{eq:gmdifferences}
g^{(m)}(\tau_1,\cdots \tau_{m-1}) = \frac{1}{I_\tx{ph}^{m}}	\tx{Tr}\left[ \mathcal{J} 	e^{\mathcal L \tau_{m-1} } 	\cdots 	\mathcal{J}  e^{\mathcal L \tau_1} \mathcal{J} \rho_\tx{NESS}\right]\, .
\end{equation}
For the effective Liouvillian of the extended Jaynes-Cummings model given in \eq{simplL}, we obtain 
\begin{equation}\label{eq:gmeffL}
g^{(m)}(\tau_1,\cdots \tau_{m-1}) = \prod_{i=1}^{m-1}\left[1-\frac{e^{-\gamma_1  \tau_i} 	\left(\gamma_2-\gamma_1 \widetilde \gm^{\tx{*(B)}}_\tx{gg}\right)}{\gamma_2}\right]\, ,
\end{equation}
where we have taken into account that $\mathcal L =\mathcal L_1$ and $\mathcal{J}= \mathcal L_1 - \mathcal L_0$. The rate coefficients are
\begin{align}
\gm_1 &= \gm_\tx{Ag} + \widetilde \gm_\tx{Ag}^{\tx{(B)}}  + \widetilde\gm^*_\tx{gA} + \widetilde \gm^{\tx{*(B)}}_\tx{gA} + \gm_\tx{gA} + \widetilde \gm^{\tx{(B)}}_\tx{gA} \, , \\ 
\gm_2 &= (	\gm_\tx{Ag} + \widetilde \gm_\tx{Ag}^{\tx{(B)}}	) (	\widetilde\gm^*_\tx{gA} + \widetilde \gm^{\tx{*(B)}}_\tx{gA} ) + (	\widetilde\gm^*_\tx{gA} + \widetilde \gm^{\tx{*(B)}}_\tx{gA} + \gm_\tx{gA} + \widetilde \gm^{\tx{(B)}}_\tx{gA}	)\, \widetilde \gm^{\tx{*(B)}}_\tx{gg}\, .
\end{align}
\subsection{Moments Expressed by the Correlation Functions}
In this subsection, we will illustrate how the $m$th factorial moment can be written as function of $g^{(m)}$. We start with the Laplace transformed factorial moments [see Eqs.~(\ref{eq:gen}) and (\ref{eq:momsm})]
\begin{equation}
M_{1,m}(s) = \braket{N^{(m)}}(s) =\int_0^\infty \partial_z^m \mathcal{M}(z,t)|_{z=\mathfrak{1}} e^{-st} \tx{d}t =  \partial^m_z \, \tx{Tr}\left[ \frac{1}{s-\mathcal{L}-z \mathcal{J}} \rho_\tx{NESS} \right]_{z=0}.
\end{equation}
We evaluate the $m$th order derivative and use $\tx{Tr}[\bullet \,  \mathcal{L}\rho_\tx{NESS} ] = \tx{Tr}[\mathcal{L} \,\bullet] =0$ to obtain
\begin{equation}
\braket{N^{(m)}}(s) = \frac{m!}{s^2} \, \tx{Tr}\left[ \left( \mathcal{J} \frac{1}{s- \mathcal{L}}\right)^{m-1}\mathcal{J} \rho_\tx{NESS}\right]  .
\end{equation}
We calculate the Laplace transform $g^{(m)}(s_1,\cdots s_{m-1})$ from \eq{gmdifferences} and immediately see the connection to the correlation function
\begin{equation} 
\braket{N^{(m)}}(s) = \frac{m!  \, I_\tx{ph}^m}{s^2}\, g^{(m)}(s_1,\cdots s_{m-1})_{s_1=\cdots s_{m-1}=s}\, .
\end{equation}
\pagebreak
\newpage
It remains to derive the inverse Laplace transform of $\braket{N^{(m)}}(s)$.
We use the relations $\mathscr{L}^{-1}[\frac{F(s)}{s}]=\int_0^t f(t_1)\tx{d} t_1$ and $\mathscr{L}^{-1}[F_\tx{a}(s) F_\tx{b}(s)]=\int_0^t f_\tx{a}(t_1)f_\tx{b}(t-t_1)\tx{d} t_1$ with $\mathscr{L}^{-1}[F(s)]=f(t)$. The factorial moments take the form
\begin{equation}
\braket{N^{(m)}}(t) = m! \int_0^t		\int_0^{t_1} \cdots{\int_0^{t_{m-1}}} 
\tx{Tr}\left[		 \mathcal{J} 	e^{\mathcal{L} t_m}	 \mathcal{J} 	e^{\mathcal{L} (t_{m-1}-t_m)}	 \mathcal{J} 	e^{\mathcal{L} (t_{m-2}-t_{m-1})} 	\cdots 	\mathcal{J} 	e^{\mathcal{L} (t_{2}-t_3)}	\mathcal{J} \rho_\tx{NESS} \right] \tx{d}t_{m}  \cdots \tx{d}t_{1}\, .
\end{equation}
We rearrange the integration domain, take equation \eq{gmcompact} into account, and obtain Eq.~(\momfromg) of the main text
\begin{equation}
\braket{N^{(m)}}(t) = m! \, I_\tx{ph}^m \int_0^t		\int_{t_1}^{t} \cdots{\int_{t_{m-1}}^{t}} 
g^{(m)}(t_1,\cdots t_{m}) \, \tx{d}t_{m}  \cdots \tx{d}t_{1}\, .
\end{equation}
The obtained relation can be simplified further if we change the integration variables $t_2, t_3, \cdots t_m$ to $\tau_1,\tau_2, \cdots \tau_{m-1}$ with $\tau_i=t_{i+1}-t_i$. Then, we rearrange the integration domain again and evaluate the integration over $t_1$ explicitly. The factorial moments read
\begin{equation}\label{eq:gmtau}
\braket{N^{(m)}}(t)=m! \, I_\tx{ph}^m \int_0^{t} \int_0^{t-\tau_1}\cdots \int_0^{t-\sum_{i=1}^{m-2}\tau_i}\left(t-\sum_{i=1}^{m-1}\tau_i \right)g^{(m)}(\tau_{1},\cdots \tau_{m-1})\,\tx{d}\tau_{m-1}\cdots\tx{d}\tau_1\, ,
\end{equation}
with $g^{(m)}(\tau_{1},\cdots \tau_{m-1})$ defined in \eq{gmdifferences}.
\subsection{First Four Moments, Correlation Functions, and Eq.~(\critone) as Function of $g^{(m)}$}
The first four factorial moments can be obtained form the first four correlation functions using \eq{gmtau} and read
\begin{align}
\braket{N^{(1)}}&=I_\tx{ph} t \,,\\
\braket{N^{(2)}}&=2! \, I_\tx{ph}^2 \int_0^t(t-\tau_1)g^{(2)}(\tau_1)\tx{d}\tau_1\, ,\\
\braket{N^{(3)}}&=3! \, I_\tx{ph}^3 \int_0^t\int_0^{t-\tau_1}(t-\tau_1-\tau_2)g^{(3)}(\tau_1,\tau_2)\tx{d}\tau_2\tx{d}\tau_1\, ,\\
\braket{N^{(4)}}&=4! \, I_\tx{ph}^4 \int_0^t\int_0^{t-\tau_1}\int_0^{t-\tau_1-\tau_2}(t-\tau_1-\tau_2-\tau_3)g^{(4)}(\tau_1,\tau_2,\tau_3)\tx{d}\tau_3\tx{d}\tau_2\tx{d}\tau_1\, .
\end{align}
The cumulants are related to the moments via Eq.~(\recursive) of the main text. We obtain for the first four cumulants
\begin{align}
C_{\tx{F},1}&=\braket{N}\, , \\
C_{\tx{F},2}&=\braket{N^{(2)}}- \braket{N}^2\, , \\
C_{\tx{F},3}&=\braket{N^{(3)}}- 3\braket{N^{(2)}}\braket{N}+2\braket{N}^3\, , \\
C_{\tx{F},4}&=\braket{N^{(4)}}- 4\braket{N^{(3)}}\braket{N}-3\braket{N^{(2)}}^2+12\braket{N^{(2)}}\braket{N}^2-6 \braket{N}^4\, ,
\end{align}
and, therefore, can relate those cumulants to the first four correlation functions.
In case of the effective Liouvillian [see \eq{simplL}], we obtain from the correlation functions given in \eq{gmeffL} the cumulants
\begin{align}
C_{\tx{F},1}&=\gm_2\gm_1^{-1}\,t\, , \\
C_{\tx{F},2}&=2\gm_2 \gm_1^{-3} (\gm_1\widetilde \gm^{\tx{*(B)}}_\tx{gg}	-	\gm_2)\, t + \mathcal{O}(t^{-1}) \, , \\
C_{\tx{F},3}&=6\gm_2 \gm_1^{-5}\left[2\gm_2^2-3 \gm_1 \gm_2 \widetilde \gm^{\tx{*(B)}}_\tx{gg}	+	\gm_1^2 \left(\widetilde \gm^{\tx{*(B)}}_\tx{gg}\right)^2\right]t	+ 	\mathcal{O}(t^{-1})\, , \\
C_{\tx{F},4}&=	24\gm_2 \gm_1^{-7} (\gm_1\widetilde \gm^{\tx{*(B)}}_\tx{gg}	-	\gm_2) \left[5\gm_2^2 - 5 \gm_1 \gm_2 \widetilde \gm^{\tx{*(B)}}_\tx{gg}	+	\gm_1^2 \left(\widetilde \gm^{\tx{*(B)}}_\tx{gg}\right)^2\right]t	 	+ 	\mathcal{O}(t^{-1})\, .
\end{align}
We recommend to derive the cumulants from measured correlation functions and then test criteria formulated as function of the cumulants [see e.g. Eqs.~(\critone), (\critthree), (\critfour), (\critthreeimproved), (\critfive), and (\critfiveimproved) of the main text]. Expressing the criteria by correlation function directly yields rather lengthy formulas as illustrated in the following for Eq.~(\critone) of the main text, i.e., $C_{\tx{F},3}=3 C_{\tx{F},2}^2 / C_{\tx{F},1}$.
First, we express the cumulants by their moments and divide by $3! I_\tx{ph}^3$. We obtain
\begin{equation}\label{eq:critermom}
\left(\frac{\braket{N^{(3)}}}{3! I_\tx{ph}^3} -\frac{t^3}{3!}\right) = \frac{2}{t} \left( \frac{\braket{N^{(2)}}}{2 I_\tx{ph}^2} -\frac{t^2}{2} \right)^2 + t \left( \frac{\braket{N^{(2)}}}{2 I_\tx{ph}^2} -\frac{t^2}{2} \right)\, ,
\end{equation}
with
\begin{align}
\frac{\braket{N^{(2)}}}{2 I_\tx{ph}^2} -\frac{t^2}{2}  &=   \int_0^t(t-\tau_1)\left[g^{(2)}(\tau_1)-1\right]\tx{d}\tau_1\, ,\\
\frac{\braket{N^{(3)}}}{3! I_\tx{ph}^3} -\frac{t^3}{3!} &= \int_0^t\int_0^{t-\tau_1}(t-\tau_1-\tau_2)\left[g^{(3)}(\tau_1,\tau_2)-1\right]\tx{d}\tau_2\tx{d}\tau_1\,.
\end{align}
We take into account that $C_{\tx{F},m}\propto t$ in the long-time limit, and thus concentrate on that part of \eq{critermom} being linear in~$t$. Terms $\propto t^{-2}$ are neglected. The criterion reads
\begin{equation}
\left(\frac{\partial_t \braket{N^{(3)}}}{3! I_\tx{ph}^3} - \frac{t^2}{2}\right) 	=	2 \left( \frac{ \partial_t \braket{N^{(2)}}}{2 I_\tx{ph}^2} -t \right)^2	+  	\left( \frac{t \, \partial_t \braket{N^{(2)}}}{2 I_\tx{ph}^2} -t^2 	+ \frac{\braket{N^{(2)}}}{2 I_\tx{ph}^2} -\frac{t^2}{2} \right)\, ,
\end{equation}
with
\begin{align}
\frac{\partial_t \braket{N^{(2)}}}{2 I_\tx{ph}^2} -t		&=	 \int_0^t \left[ g^{(2)}(\tau_1) -1 \right]\tx{d}\tau_1\, ,\\
\frac{\partial_t \braket{N^{(3)}}}{3! I_\tx{ph}^3} - \frac{t^2}{2}	&	=\int_0^t\int_0^{t-\tau_1}\left[g^{(3)}(\tau_1,\tau_2)-1\right]\tx{d}\tau_2\tx{d}\tau_1\,.
\end{align}
Finally, we can rewrite Eq.~(\critone) in the form
\begin{equation}
\lim_{t\to \infty} \int_0^t\int_0^{t-\tau_1}\left[g^{(3)}(\tau_1,\tau_2)-1\right]\tx{d}\tau_2\tx{d}\tau_1\\
=\lim_{t\to \infty} \left\{ 2 \left( \int_0^t \left[g^{(2)}(\tau_1)-1\right]\tx{d}\tau_1\right)^2+\int_0^t (2t-\tau_1)\left[g^{(2)}(\tau_1)-1\right]\tx{d}\tau_1\, \right\}.
\end{equation}

\section{APPLICATION OF EQ.~(\critthree) AND EQ.~(\critfour)}\label{app:signcrit}
The sign criteria given in Eq.~(\critthree) and (\critfour) of the main text hold in case of a single excited state. These criteria are not necessarily fulfilled in case of an additional state B. In contrast, we obtain from \eq{simplL} opposite signs
\begin{align}
&\frac{C_{\tx{F},3}-\widetilde{C}_{3}}{C_{\tx{F},3}}=\frac{1}{1+2c}+\mathcal{O}(\Delta t)>0 \, , \\
&\frac{w-\widetilde{w}}{w}= - t \frac{(c+1)c\, (\eta \, \gm^{\tx{*(B)}}_\tx{gg})^3 }{\gamma_\text{out}^2}+\mathcal{O}(\Delta t)< 0 \, ,
\end{align}
if the sampling time $\Delta t$ is smaller than a certain critical value, which for the parameters used in Fig.~3 of the main text is about $17\,\tx{ns}$. Here, the dimensionless coefficient is $c= (\gm_{\tx{Ag}}+\widetilde \gm_\tx{Ag}^{\tx{(B)}})(\widetilde\gm^*_\tx{gA}+\widetilde\gm^{\tx{*(B)}}_\tx{gA})/(\gm^{\tx{*(B)}}_\tx{gg} \gm_\tx{out})$ with $\gm_\tx{out}=\gm_{\tx{gA}}+\widetilde \gm_\tx{gA}^{\tx{(B)}}+\widetilde\gm^*_\tx{gA}+\widetilde\gm^{\tx{*(B)}}_\tx{gA}$.

\section{CONTINUOUS ERROR BARS}\label{app:bar}
A photon stream is recorded in experiment only during a finite time span $T$. Consequently, statistical quantities can be calculated only from a finite amount of data. To model the resulting stochastic error, we apply the procedure introduced in Ref.~\onlinecite{kleinherbers_pushing_2022}. First, we express the considered quantity $g$ by generalized factorial moments $M_{s,m}=\sum_n f_{s,m,N} P_N(t)$ with $f_{s,m,N}=N^{(m)}s^{N-m}$. For example, the quantity $(C_{\tx{F},3}-\widetilde C_{\tx{F},3})/C_{\tx{F},3}$ discussed in the main text can be written as function of $M_{1,1},M_{1,2}$, and $M_{1,3}$. Then, we calculate
\begin{equation}
\tx{RMSE}^2= \frac{t}{T} \sum_N P_N(1-P_N)\left[\sum_\mathbf{k} \frac{\partial g}{\partial M_{\mathbf{k}}}\left(f_{\mathbf{k},N} - M_{\mathbf{k}}\right)\right]^2,
\end{equation}
where the latter sum runs over the indices $\mathbf{k}=(s,m)$ of the involved moments. Here, RMSE is the root-mean square error. We obtain continuous error bars of the form $g\pm \tx{RMSE}$. In Fig.~4 of the main text, three different functions $g(\Delta t)$, $g(2\Delta t)$, and $g(3\Delta t)$ are involved, each with its own probability distribution $P_N$. We calculate the error bars from the RMSE of $g(\Delta t)$. We have verified the applicability of this simplification by simulating actual photon streams.

\pagebreak
\section{GENERAL TWO-STATE MODEL}\label{app:genmodel}
\begin{figure*}[t]
	\includegraphics[width=0.45\textwidth]{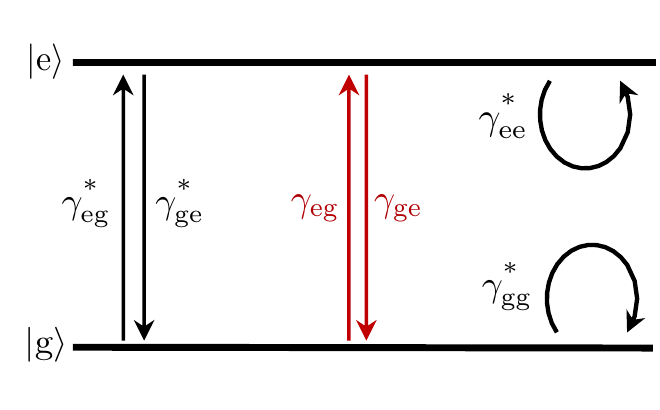}
	\caption{General form of an incoherent model describing the dynamics between two states $\ket{\text{g}}$ and $\ket{\text{e}}$. Counted transitions are depicted by black arrows and not counted transitions by red arrows.}\label{fig:fig5}
\end{figure*}
In this section, we elucidate what kind of information one can obtain from the inverse counting statistics procedure in case of a two dimensional Liouvillian, i.e., except the two states $\ket{\text{g}}$ and $\ket{\text{e}}$, all other states are transient and their occupation probability vanishes. The Liouvillian takes the general form
\begin{equation}
\mathcal{L}_{z} =\!
\begin{pmatrix}
(z\tx{-}1)\, \gm^{\tx{*}}_\tx{gg}\tx{\,-\,} \gm^{\tx{*}}_\tx{eg}\tx{\,-\,} \gm_\tx{eg} & z\, \gm^{\tx{*}}_\tx{ge}\tx{\,+\,} \gm_\tx{ge}\\
z\, \gm^{\tx{*}}_\tx{eg}\tx{\,+\,} \gm_\tx{eg}  & (z\tx{-}1)\, \gm^{\tx{*}}_\tx{ee}\tx{\,-\,} \gm^{\tx{*}}_\tx{ge}\tx{\,-\,} \gm_\tx{ge}
\end{pmatrix}
\end{equation}
in the basis $(\rho_{\ket{\tx{g}}},\rho_{\ket{\tx{e}}} )^T$ and is illustrated in~\fig{fig5}. The self-loop transitions at rate $\gm^{\tx{*}}_\tx{gg}$ and $\gm^{\tx{*}}_\tx{ee}$ increase the counter but do not exchange occupation probability between $\ket{\tx{g}}$ and $\ket{\tx{e}}$. These transitions can occur due to additional hidden states or the presence of background photons.
The characteristic polynomial reads
\begin{equation}\label{eq:charpolgenapp}
\chi (z,\lambda)=\lambda^2+(a_{01}+a_{11}z) \lambda+ a_{00}+a_{10} z +a_{20} z^2\, .
\end{equation}
Since $\lambda=0$ is an eigenvalue for $z=1$, we have $a_{10}=-a_{00}-a_{20}$. The other coefficients are 
\begin{align}
a_{20}&=\gm^{\tx{*}}_\tx{ee}\gm^{\tx{*}}_\tx{gg} - \gm^{\tx{*}}_\tx{ge} \gm^{\tx{*}}_\tx{eg}, \label{eq:appa20rate} \\
a_{11}&=-\gm^{\tx{*}}_\tx{ee} -\gm^{\tx{*}}_\tx{gg} , \label{eq:appa11rate} \\
a_{01}&= \gm_\tx{ge} + \gm_\tx{eg} + \gm^{\tx{*}}_\tx{ge} + \gm^{\tx{*}}_\tx{eg} + \gm^{\tx{*}}_\tx{gg} + \gm^{\tx{*}}_\tx{ee}, \\
a_{00}&=\left(\gm_\tx{ge}+\gm^{\tx{*}}_\tx{ge}+\gm^{\tx{*}}_\tx{ee} \right) \left(\gm_\tx{eg}+\gm^{\tx{*}}_\tx{eg}+\gm^{\tx{*}}_\tx{gg} \right) -\gm_\tx{ge} \gm_\tx{eg} . \label{eq:app00rate}
\end{align}
They can be obtained from the first four cumulants. We form a system of the linear equations from the first four derivatives $\partial_z^m \chi(z,\lambda_\tx{max}(z))_{z=1}=0$, use Eq.~(\lamax) of the main text to replace the derivatives of the eigenvalues by cumulants, and solve for the coefficients.
We obtain
\begin{align}
a_{20}&=\lim_{t\to \infty}\left(\frac{9C_{\tx{F},2}^4-6 C_{\tx{F},3} C_{\tx{F},2}^2C_{\tx{F},1}}{(3C_{\tx{F},4}C_{\tx{F},2}-4C_{\tx{F},3}^2)t^2} +\frac{C_{\tx{F},1}^2}{t^2}  \right)   ,  \label{eq:appa20}\\
a_{11}&=\lim_{t\to \infty}\left( \frac{6 C_{\tx{F},3} C_{\tx{F},2}^2}{(3C_{\tx{F},4}C_{\tx{F},2}-4C_{\tx{F},3}^2)t} -\frac{2C_{\tx{F},1}}{t} \right)  , \label{eq:appa11} \\
a_{01}&=\lim_{t\to \infty}\left( -\frac{6 (C_{\tx{F},3}+3C_{\tx{F},2}) C_{\tx{F},2}^2}{(3C_{\tx{F},4}C_{\tx{F},2}-4C_{\tx{F},3}^2)t} +\frac{2C_{\tx{F},1}}{t} \right)   ,  \label{eq:appa01}\\
a_{00}&=\lim_{t\to \infty}\left(\frac{9C_{\tx{F},2}^4-6 (C_{\tx{F},3}+3C_{\tx{F},2}) C_{\tx{F},2}^2C_{\tx{F},1}}{(3C_{\tx{F},4}C_{\tx{F},2}-4C_{\tx{F},3}^2)t^2} +\frac{C_{\tx{F},1}^2}{t^2}  \right)   .  \label{eq:appa00}
\end{align}
Both coefficients $a_{00}$ and $a_{01}$ are always larger than zero and do not give much physical insight. In contrast, a vanishing coefficient $a_{11}$ reveals the absence of self-loop transitions ($\gm^{\tx{*}}_\tx{ee} = \gm^{\tx{*}}_\tx{gg}=0$). A vanishing coefficient $a_{20}$ reveals the absence of two counted self-loop ($\gm^{\tx{*}}_\tx{ee}\gm^{\tx{*}}_\tx{gg}=0$) and two counted ordinary transitions ($\gm^{\tx{*}}_\tx{ge} \gm^{\tx{*}}_\tx{eg}=0$). In contrast to $a_{11}$, the coefficient is vanishing even in case of a single nonvanishing self-loop transition. From $a_{11}=0$ and $a_{20}=0$, we find that the fourth cumulant must fulfill in the long-time limit
\begin{align}
\widetilde{C}_{4}&=\frac{4}{3}\frac{C_{\tx{F},3}^2}{C_{\tx{F},2}} + \frac{C_{\tx{F},3}C_{\tx{F},2}}{C_{\tx{F},1}}\, , \label{eq:appC4a}\\
\widetilde{C}_{4}&=\frac{4}{3}\frac{C_{\tx{F},3}^2}{C_{\tx{F},2}} + 2\frac{C_{\tx{F},3}C_{\tx{F},2}}{C_{\tx{F},1}}-3\frac{C_{\tx{F},2}^3}{C_{\tx{F},1}^2}\, , \label{eq:appC4b}
\end{align}
respectively. The latter condition is Eq.~(\critfive) in the main text. If only one ordinary transition is counted, both coefficients are vanishing $a_{11}=a_{20}=0$. Thus, we can combine \eq{appC4a} and \eq{appC4b} to recover Eq.~(\critone).

\addcontentsline{toc}{part}{References}
\stoptocentries

\end{document}